`

# A Dual Physics-Informed Kolmogorov–Arnold Neural Network Framework for Continuum Topology Optimization

Junyuan Zhang [a], Jing Cao [b], Abdullah Dawar [a], Kun Cai [a*], Qinghua Qin [c*]

[a] *School of Science, Harbin Institute of Technology, Shenzhen 518055, China*

[b] *State Key Laboratory of Water Engineering and Ecological Environment in Arid Areas, Xi'an Institute of Technology, Xi'an 710048, China*

[c] *Institute of Advanced Interdisciplinary Technology, Shenzhen MSU-BIT University, Shenzhen, 518172, China*

**Abstract**: In continuum topology optimization (TO), two essential procedures are involved: structural analysis through the solution of partial differential equations (PDEs) and the subsequent update of design variables. Both procedures can be addressed by training neural networks using the corresponding physical information. Accordingly, Physics-Informed Neural Network (PINN)–based algorithms have been developed for TO. However, PINN-based methods suffer from several notable limitations, including high computational cost, spectral bias, and limited adaptability in solving PDEs. To overcome these challenges, this study proposes a novel algorithm that incorporates two Higher-Order ReLU–based Kolmogorov–Arnold Networks (HRKANs). Specifically, a displacement-informed HRKAN (d-HRKAN) is designed to predict PDE solutions, while a sensitivity-informed HRKAN (s-HRKAN) is developed to perform sensitivity analysis for updating design variables. For convenience, the proposed approach is referred to as the Dual Physics-Informed Kolmogorov–Arnold Networks–based Topology Optimization (DPIKAN-TO) method. By leveraging learnable activation functions, the proposed neural networks can accurately approximate the responses of complex structural systems. Moreover, compared with conventional PINN-based methods, DPIKAN-TO demonstrates significantly improved computational efficiency and reduced computational cost. Numerical examples show that DPIKAN-TO can successfully identify optimal material layouts for linear structures, compliant mechanisms, and fluid–solid coupled systems. Furthermore, owing to the use of learnable activation functions, the proposed framework can be readily extended to structural optimization problems governed by new types of PDEs.

**Keywords**: Topology optimization; Physics-Informed Neural Networks; Kolmogorov-Arnold Networks; Partial Differential Equations; Sensitivity analysis

## 1. Introduction

Continuum topology optimization (TO) is a fundamental methodology in structural optimization that seeks to improve specified performance objectives through the optimal distribution of material within a given design domain (Martin P Bendsøe 2008). In contrast to conventional size and shape optimization techniques, topology optimization offers substantially greater design freedom by allowing variations in the structural topology itself, thereby enabling more efficient material utilization. After nearly four decades of development, topology optimization theory and methods have reached a high level of maturity. A wide range of algorithms have been proposed and successfully applied, including the Solid Isotropic Material with Penalization (SIMP) method (Martin P Bendsøe 1989;Sigmund 2001), level-set methods (LSM) (Van Dijk et al. 2013), Evolutionary Structural Optimization (ESO) (Huang and Xie 2010), Material Field Series Expansion (MFSE) (Luo and Bao 2019), phase field methods (PFM) (Takezawa et al. 2010), the normalized field product (NFP) approach for parameter-free density evaluation(Singh et al. 2025), and integer linear programming (ILP) formulations for strictly binary topology optimization(Sivapuram and Picelli 2018). These approaches have been widely adopted across diverse engineering disciplines, including aerospace, automotive, and civil engineering (Cavazzuti et al. 2011).

However, continuum topology optimization typically involves a large number of design variables, leading to an exponential increase in computational cost. To improve computational efficiency, researchers have explored solutions from multiple perspectives. These efforts include: (1) optimizing computational architectures by developing acceleration strategies based on parallel computing (Aage et al. 2015); (2) enhancing numerical algorithms through multigrid-assisted analysis methods (Yin et al. 2022); (3) advancing model order reduction techniques, such as surrogate modeling and subspace projection methods (Gogu 2015;Amir et al. 2012;Zheng et al. 2020); and (4) reducing the dimensionality of design variables, for example, via the Moving Morphable Component (MMC) method (Guo et al. 2016). These approaches have demonstrated significant acceleration for specific topology optimization problems. Nevertheless, challenges related to scalability, strongly nonlinear mechanical behavior, and multiphysics coupling persist, underscoring the need for new algorithms that further extend the applicability of topology optimization to complex engineering designs.

In recent years, the rapid advancement of deep learning has attracted considerable attention to the integration of neural networks with topology optimization. Early studies primarily focused on data-driven surrogate (proxy) modeling approaches, in which pre-trained neural networks learn mapping relationships from large datasets to enable fast predictions, thereby serving as efficient alternatives to conventional topology optimization methods. For instance, Sosnovik and Oseledets (2019) proposed a convolutional neural network (CNN)-based U-Net architecture that reformulates topology optimization as an image segmentation problem. This method learns the mapping between structural topology and the corresponding objective function, significantly reducing computational cost. Similarly, Rawat and Shen (2018) introduced a topology optimization framework that combines generative adversarial networks (GANs) with CNNs. In their approach, GANs are employed for topology generation, while CNNs are used for performance prediction, followed by structural validation, enabling rapid and reliable topology optimization.

However, data-driven approaches exhibit inherent limitations in topology optimization (Shin et al. 2023). First, constructing sufficiently large labeled datasets for model training entails substantial computational cost. Second, in the absence of explicit physical constraints during training, the generated designs often fail to satisfy complex boundary conditions. In scenarios involving blurred boundaries and irregular geometries, the reliability and effectiveness of such designs are therefore difficult to ensure (Zhang et al. 2021). To overcome these challenges, recent research has increasingly focused on physics-informed methodologies, particularly Physics-Informed Neural Networks (PINNs). Introduced by Raissi et al. (2019) ), PINNs embed physical knowledge directly into the training process by incorporating governing partial differential equations (PDEs) into neural network formulations. Building on this concept, Chandrasekhar and Suresh (2021), proposed the Topology Optimization using Neural Networks (TOuNN) framework, which integrates the objective and constraint functions of the optimization problem into a neural network representation of the structural density field. This formulation enables the updating of design variables through structural analysis coupled with neural network backpropagation. Subsequently, He et al.(2023b) developed a topology optimization framework based on the Deep Energy Method (DEM), wherein a DEM-based PINN replaces conventional structural analysis to obtain the displacement field, while optimization is performed using sensitivity analysis and the optimality criteria method. Extending this line of research, Jeong et al. (2023) introduced the Complete Physics-Informed Neural Network-based Topology Optimization (CPINNTO) framework. By coupling a DEM-PINN with a sensitivity-analysis PINN (S-PINN), this approach eliminates the need for traditional finite element and sensitivity analyses, thereby establishing a fully PINN-driven topology optimization paradigm. Moreover, the framework has been successfully extended to address advanced topology optimization problems, including those involving geometric nonlinearity and multiscale structures (Jeong et al. 2025).

To date, PINN-based approaches have been extended to address increasingly challenging topology optimization problems, including geometric and material nonlinearities as well as multi-material design (Jeong et al. 2025;Zhang et al. 2021). Nevertheless, the application of neural networks to problems involving complex loading conditions and multiphysics coupling remains relatively limited. Moreover, most existing studies rely predominantly on data-driven strategies (Chadha and Kumar 2025), with insufficient explicit incorporation of multiphysics knowledge to guide and enhance the optimization process. In addition, conventional PINN architectures typically employ fixed activation functions, which restrict their capacity to accurately approximate complex structural responses and capture high-frequency features(Farea and Celebi 2025).

Despite these advances, most existing neural-network-based topology optimization frameworks still follow the conventional computational paradigm in which equilibrium analysis and topology evolution are treated as separate iterative procedures. Fundamentally, continuum topology optimization involves two strongly coupled physical processes: the PDE-governed evolution of the structural equilibrium field and the sensitivity-driven evolution of the material distribution field. Existing approaches typically handle these two processes independently through finite element analysis and gradient-based optimization updates, respectively. Consequently, the overall optimization procedure remains computationally fragmented and difficult to generalize toward strongly coupled multiphysics systems.

More importantly, topology optimization is intrinsically associated with localized high-frequency field evolution. During the iterative evolution of material layouts, the density field progressively approaches a nearly binary distribution, which leads to sharp solid–void interfaces, localized stiffness redistribution, and rapidly varying sensitivity fields. Therefore, neural-network-based topology optimization requires not only the ability to approximate smooth displacement fields, but also the capacity to represent sharp topology interfaces and localized sensitivity variations. This intrinsic characteristic partly explains why conventional MLP-based PINN architectures may produce oversmoothed material layouts, blurred structural boundaries, and reduced capability in representing slender load-bearing members.

The objective of the present work is therefore not merely to improve neural network approximation accuracy, but to establish a dual physics-informed surrogate framework capable of simultaneously representing equilibrium evolution and topology evolution within a unified differentiable optimization paradigm.

Within such a framework, the capability of the neural representation to capture sharp topology interfaces, localized high-frequency features, and strongly nonlinear sensitivity evolution becomes critically important. To address these limitations, recent developments in Kolmogorov–Arnold Networks (KANs) have introduced learnable activation representations with improved approximation flexibility. In particular, the Higher-order ReLU-based KAN (HRKAN) demonstrates enhanced capability for representing localized high-frequency structural and sensitivity features while maintaining computational efficiency and differentiability, making it particularly suitable for coupled surrogate modeling in topology optimization.

To enhance computational efficiency and expand the applicability of neural network-based topology optimization in multiphysics environments, this study integrates HRKAN with DEM-PINN and S-PINN frameworks. Two novel modules are introduced: the displacement-informed HRKAN (d-HRKAN) and the sensitivity-informed HRKAN (s-HRKAN), which are designed to replace conventional structural analysis and sensitivity-based design updates, respectively, The key contribution of this work is not merely replacing MLPs with HRKANs, but establishing a dual physics-informed surrogate framework in which d-HRKAN approximates PDE-governed equilibrium states, while s-HRKAN approximates the implicit optimization landscape and sensitivity evolution. This enables topology optimization to be reformulated as a coupled neural field evolution problem rather than a conventional FEM-driven iterative procedure. Building upon these components, the proposed Dual Physics-Informed Kolmogorov–Arnold Network-based Topology Optimization (DPIKAN-TO) framework demonstrates enhanced versatility owing to its ability to accommodate a broad class of governing PDEs. Compared with existing PINN-based topology optimization approaches, the proposed dual-surrogate strategy significantly reduces the computational cost of the optimization loop and offers a more efficient paradigm for addressing complex multiphysics topology optimization problems.

The remainder of this article is organized as follows. Section 2 reviews continuum topology optimization methods and introduces the DPIKAN-TO methodology, establishing the theoretical foundation of the proposed approach. Section 3 presents the implementation details and workflow of the DPIKAN-TO framework integrated with HRKAN, with particular emphasis on the architectures of the proposed d-HRKAN and s-HRKAN modules and their roles within the overall optimization process. Section 4 evaluates the feasibility and performance of DPIKAN-TO through a series of numerical examples, including two-dimensional (2D) compliance minimization, rigid structure and compliant mechanism design under fluid pressure loading, and three-dimensional (3D) compliance minimization. Finally, Section 5 summarizes the advantages and limitations of the proposed DPIKAN-TO approach.

`

## 2. Methodology

This section introduces the fundamental mathematical model of continuum topology optimization and the theoretical foundations of KAN. First, the standard formulation of the optimization problem is reviewed. Next, the core concepts of KANs are presented. Finally, the proposed DPIKAN-TO framework, which integrates these ideas, is described.

### *2.1 Continuous topology optimization theory*

Topology optimization is a computational technique that determines the optimal material distribution within a prescribed design domain. The following subsection presents the mathematical formulation that formally defines the general problem.

#### *2.1.1 Mathematical formulation of optimization problems*

In continuum topology optimization, the objective is to determine the optimal distribution of structural material within a prescribed design domain Ω by introducing a material density function $\rho(x)\in[0,\ 1]$. The optimization procedure generally involves two fundamental steps: (i) structural analysis to evaluate the system response and (ii) iterative updating of the design variables. For the compliance minimization problem, the topology optimization model can be formulated as follows:

$$
\begin{aligned}
&\text{Find } \boldsymbol{\rho}=\left[\rho_1,\rho_2,\ldots,\rho_n\right]\\
&\min\ C\left(\boldsymbol{\rho},\boldsymbol{u}\left(\boldsymbol{\rho}\right)\right)\\
&\text{s.t. } \mathrm{R}\left(\boldsymbol{\rho},\boldsymbol{u}\left(\boldsymbol{\rho}\right)\right)=0\quad\left(\text{PDEs}\right)\\
&\qquad g_i\left(\boldsymbol{\rho}\right)\le 0\qquad i=1,2,\ldots\\
&\qquad h_j\left(\boldsymbol{\rho}\right)=0\qquad j=1,2,\ldots
\end{aligned}
\tag{1}
$$

where, $C$ denotes the objective function, namely the structural compliance, with lower values corresponding to higher structural stiffness. The vector $\boldsymbol{u}$ represents the structural displacement field, and $\rho_i$ denotes the pseudo-density associated with the $i$th material point or finite element. The term R represents the residuals of the governing physical equations, corresponding to the PDE constraints that must be satisfied in structural analysis. Finally, $g_i(\boldsymbol{\rho})$ and $h_i(\boldsymbol{\rho})$ denote the inequality and equality constraints of the optimization problem, respectively.

### 2.1.2 Density–Stiffness interpolation model

The essence of continuum topology optimization lies in determine the optimal material distribution within a design domain through iterative updates of the material layout. Because variations in material distribution directly affect the local stiffness, density–stiffness interpolation schemes are commonly adopted to model this relationship. Among these schemes, the Solid Isotropic Material with Penalization (SIMP) approach is the most widely used. For a discretized structur analyzed using finite element method, the Young's modulus of the material in element *i*, denoted by $E(\rho_i)$, can be expressed as follows (Martin P Bendsøe and Sigmund 1999):

$$E\left(\rho_i\right) = E_{\min} + \rho_i^p \left(E_{\text{solid}} - E_{\min}\right), \quad (2)$$

where, $E_{\text{solid}}$ denotes the Young's modulus of the solid material. To prevent singularities in the global stiffness matrix, a small stiffness value, denoted as $E_{\min}$, is introduced for the void region to ensure stable numerical computations. The parameter $p$, serving as a penalty factor, is employed to suppress intermediate density values, thereby driving the design variables toward binary values close to 0 and 1, thus producing a sharp boundary between solid and void. The choice of the penalty factor $p$ has a direct impact on the clarity of the final topology boundary. The value of $p$ that is too small may result in an indistinct structural boundary, whereas an excessively large $p$ can lead to numerical convergence difficulties and additional numerical issues. In practice, a value of $p$=3 is commonly adopted. The nonlinear density–stiffness interpolation in Eq.(2) also explains the localized high-frequency characteristics of topology optimization. As the density field approaches a nearly binary distribution, sharp solid–void interfaces induce local stiffness redistribution and rapidly varying sensitivity responses. Thus, topology optimization can be regarded as a coupled field-evolution process involving density evolution, equilibrium redistribution, and sensitivity propagation.

## 2.2 KAN theory

### 2.2.1 KAN Network Architecture

Inspired by the KAN representation theorem (Schmidt-Hieber 2021), the KAN proposed by Liu et al. (2024) distinguishes itself from traditional multi-layer perceptions (MLPs) by not employing a fixed activation function on its nodes. Instead, it utilizes a learnable activation function $\varphi(\boldsymbol{x})$, which is derived from a combination of basis functions and bias functions that are formulated as linear combinations of B-spline curves on the edges. Consequently, the parameter space of KAN is characterized by both the weight matrices and the activation functions. This configuration allows the trained network to achieve optimality not only in terms of the weight matrices but also regarding the functional representation of the activations:

$$\phi(x) = \omega_{\mathrm{s}} \sum_{i=1}^{G+k} c_i B_i(x) + \omega_{\mathrm{b}} \frac{x}{\left(1+e^{-x}\right)}, \tag{3}$$

where, $\omega_{\mathrm{s}}$ and $\omega_{\mathrm{b}}$ denote the trainable weight parameters of the basis and bias functions, while $c_i$ represents the trainable weight parameter linked to the corresponding B-spline curve. The parameter $G$ specifies the number of grid points arranged within the support interval of the basis function, thereby controlling fitting accuracy and the network's capacity to capture high-frequency features. The order $k$ determines the smoothness of the basis function and, consequently, the continuity of higher-order derivatives in the learned mapping. The network structure of KAN can be represented by an integer array, i.e.,

$$\mathrm{NN} = \left[d_1,\ d_2,\ \dots,\ d_{n-1},\ d_n\right], \tag{4}$$

where $d_1$ and $d_n$ represent the neuron counts of the input and output layers, respectively. The mid-term $d_i$ denotes the neuron count of an intermediate layer ($i$=2, 3, …, $n$-1). Consequently, the activation function between layers ($l$, $l$+1) in KAN can be expressed in matrix form as:

$$\boldsymbol{x}_{l+1} = \mathrm{KAN}_{l,l+1}(\boldsymbol{x}_l) = \begin{pmatrix} \phi_{1,1}(\cdot) & \phi_{1,2}(\cdot) & \cdots & \phi_{1,m}(\cdot) \\ \phi_{2,1}(\cdot) & \phi_{2,2}(\cdot) & \cdots & \phi_{2,m}(\cdot) \\ \vdots & \vdots & \ddots & \vdots \\ \phi_{q,1}(\cdot) & \phi_{q,2}(\cdot) & \cdots & \phi_{q,m}(\cdot) \end{pmatrix} \boldsymbol{x}_l, \tag{5}$$

where, $\boldsymbol{x}_l$ and $\boldsymbol{x}_{l+1}$ denote the activations of the $l$-th and ($l$+1)-th layers in KAN ($l$ = 1, 2, …, $n$-1). The function $\phi_{j,i}(\cdot)$ is the learnable activation mapping from neuron $i$ in layer $l$ to $j$ in layer ($l$+1). $m$ and $q$ denote the neuron counts of the respective layers, determining the size of the activation matrix.

### *2.2.2 HRKAN Network Architecture*

One major drawback of KANs is that they are unable to scale to larger problems due to the higher computing cost associated with the complexity of B-spline fundamental functions. Although a number of effective basis functions have been proposed recently as alternatives to B-splines in KANs, these approaches frequently compromise the smoothness necessary for simulating intricate data patterns. The HRKAN framework uses a new and simplified Higher-order ReLU (H-ReLU) activation function to handle this trade-off, which is a significant architectural breakthrough. In addition to provably preserving the crucial continuity of higher-order derivatives necessary for precise gradient flow and function approximation, this tactical replacement greatly improves compute efficiency and scalability during training and inference (So and Yung 2025). Thus, HRKAN qualifies for use in more challenging contexts by achieving a more promising computational track without sacrificing the expressive capacity that characterizes the KAN paradigm.

$$r_i(x) = \left(\mathrm{ReLU}(e_i - x) \times \mathrm{ReLU}(x - s_i)\right)^{\beta} \times c_{\beta}, \tag{6}$$

where $e_i$ and $s_i$ denote the support interval points of the trainable basis functions. Assuming that the function approximation interval is $x \in [-1,1]$, with $G$ grid points and polynomial order $K$, the endpoints of the support interval of the basis function $r_i(x)$ are initialized as follows: $s_i = 2\frac{i-k-1}{G} - 1,\ e_i = 2\frac{i}{G} - 1$. $c_{\beta} = \left(\frac{2}{e_i - s_i}\right)^{2\beta}$ is a normalization parameter introduced to guarantee that each basis function has a unit height, while $r_i(x)$ denotes the basis function supported on the interval $[s_i, e_i]$. Accordingly, the activation function is expressed as:

$$\phi(x) = \sum_{i=1}^{G+k} \omega_i r_i(x) \tag{7}$$

where $\omega_i$ is the trainable weight of the basis function $r_i(x)$.

## 3. DPIKAN-TO framework: Architecture and Implementation

This section presents the core mechanisms and operational mechanics of the projected DPIKAN-TO framework. Initially, the network architectures of the two dedicated submodules, d-HRKAN for high fidelity structural analysis and s-HRKAN for effective design variable updates, within the HRKAN-integrated framework. Further, the computational workflow and implementation details are elaborated subsequently.

### *3.1 The architecture of d-HRKAN*

The fundamental essence of neural network training lies in optimizing the network parameters to minimize a predefined loss function. Consequently, employing neural networks to solve PDEs can be viewed as an optimization problem concerning the network parameters. For the d-HRKAN architecture, the critical aspect lies in incorporating the structural potential energy into the formulation of the loss function. This transforms the PDE into a variational optimization problem, where the network learns that solution by minimizing the total potential energy of the system. By reducing the system's total potential energy, the network learns the solution to the PDEs, which is now a variational optimization problem.

For isotropic linear elastic structures under the small deformation assumption, the strain and stress can be expressed in terms of the displacement field $\boldsymbol{u}$ and the Lamé parameters $\lambda$ and $\mu$:

$$\boldsymbol{\varepsilon}=\frac{1}{2}\left(\nabla\boldsymbol{u}+\nabla\boldsymbol{u}^{\mathrm{T}}\right),\quad \boldsymbol{\sigma}=\boldsymbol{D}:\boldsymbol{\varepsilon}=\lambda\operatorname{tr}\left(\boldsymbol{\varepsilon}\right)+2\mu\boldsymbol{\varepsilon}, \tag{8}$$

where, $\boldsymbol{\varepsilon}$, $\boldsymbol{\sigma}$, and $\boldsymbol{D}$ represent the strain tensor, stress tensor, and elasticity tensor, respectively. For plane stress conditions, the Lamé constants are given by:

$$\lambda=\frac{Ev}{(1+v)(1-v)},\quad \mu=\frac{E}{2(1+v)}, \tag{9}$$

where, $E$ and $v$ denote the Young’s modulus and Poisson’s ratio of the material. Accordingly, the elastic strain energy of the structure can be expressed as:

$$U\left(\boldsymbol{u}\right)=\frac{1}{2}\int_{\Omega}\boldsymbol{\varepsilon}\left(\boldsymbol{u}\right):\boldsymbol{D}:\boldsymbol{\varepsilon}\left(\boldsymbol{u}\right)\mathrm{d}\Omega \tag{10}$$

In the absence of body forces and under the Neumann boundary condition $\Gamma_{\boldsymbol{t}}$ and Dirichlet boundary condition $\Gamma_{\boldsymbol{u}}$, the external work done by the structural loads can be expressed as:

$$W(\boldsymbol{u}) = \frac{1}{2}\int_{\Gamma_t} \bar{\boldsymbol{t}} \cdot \boldsymbol{u} \mathrm{d}\Gamma \tag{11}$$

where $\bar{\boldsymbol{t}}$ is the external traction applied on the boundary $\Gamma_{\boldsymbol{t}}$. The total potential energy of the structure is thus given by:

$$\Pi(\boldsymbol{u}) = U(\boldsymbol{u}) - W(\boldsymbol{u}). \tag{12}$$

By incorporating the HRKAN into the DEM-PINN framework, we derive d-HRKAN as shown in Fig. 1. In 2D problems, the network utilizes planar coordinates ($x$, $y$) as input and produces the corresponding nodal degrees of freedom (DOF), specifically the displacements ($u$, $v$):

$$\begin{aligned} u(x, y) &= \mathrm{KAN}(x, y; \boldsymbol{\theta}), \\ v(x, y) &= \mathrm{KAN}(x, y; \boldsymbol{\theta}), \end{aligned} \tag{13}$$

where $\boldsymbol{\theta}$ represents the trainable parameters of the network. In topology optimization, accurately solving the displacement field is essential, as both the objective and constraint sensitivities explicitly depend on it. Therefore, within d-HRKAN framework, this requires precise modeling of the strain energy term, which directly affects the accuracy of displacement predictions and the convergence of the optimized topology. To this end, the element strain energy is computed using shape functions and Gaussian quadrature (He et al. 2023a).

It is important to highlight that d-HRKAN does not include an explicit boundary condition loss term within its loss function. Instead, Dirichlet boundary conditions are enforced through hard constraints on the output displacements (Lu et al. 2021). Consequently, the loss function of d-HRKAN can be expressed as:

$$\arg\min L(\boldsymbol{\theta}) = \Pi(\boldsymbol{u}; \boldsymbol{x}, \boldsymbol{\theta}) = \frac{1}{2}\sum_{i\in\Omega} \omega_i \boldsymbol{\varepsilon}_\theta(\boldsymbol{x}_i) : \boldsymbol{D} : \boldsymbol{\varepsilon}_\theta(\boldsymbol{x}_i) - \sum_{j\in\Gamma_t} \bar{\boldsymbol{t}} \cdot \boldsymbol{u}_\theta(\boldsymbol{x}_j), \tag{14}$$

where, $\boldsymbol{x}_i$ denotes the spatial coordinates of the $i$-th integration point, $\omega_i$ is the corresponding Gaussian weight, while $\boldsymbol{u}_\theta$ and $\boldsymbol{t}$ represent the displacement and load vectors at the integration point. Since each basis function in the KAN network is defined over finite support intervals (typically defined as [0,1] or [-1,1]), the spatial coordinates ($x$, $y$) of the sampled collocation points must be normalized accordingly. The normalization of the network inputs can be expressed as:

$$\boldsymbol{x}^* = 2\times\frac{\boldsymbol{x}-x_{\min}}{x_{\max}-x_{\min}}-1 \in [-1,\ 1]. \tag{15}$$

Moreover, to ensure that the inputs remain within the valid support of the basis functions throughout the network, batch normalization is applied to the inputs and outputs of each hidden layer, as well as to the input of the output layer (Ioffe and Szegedy 2015).

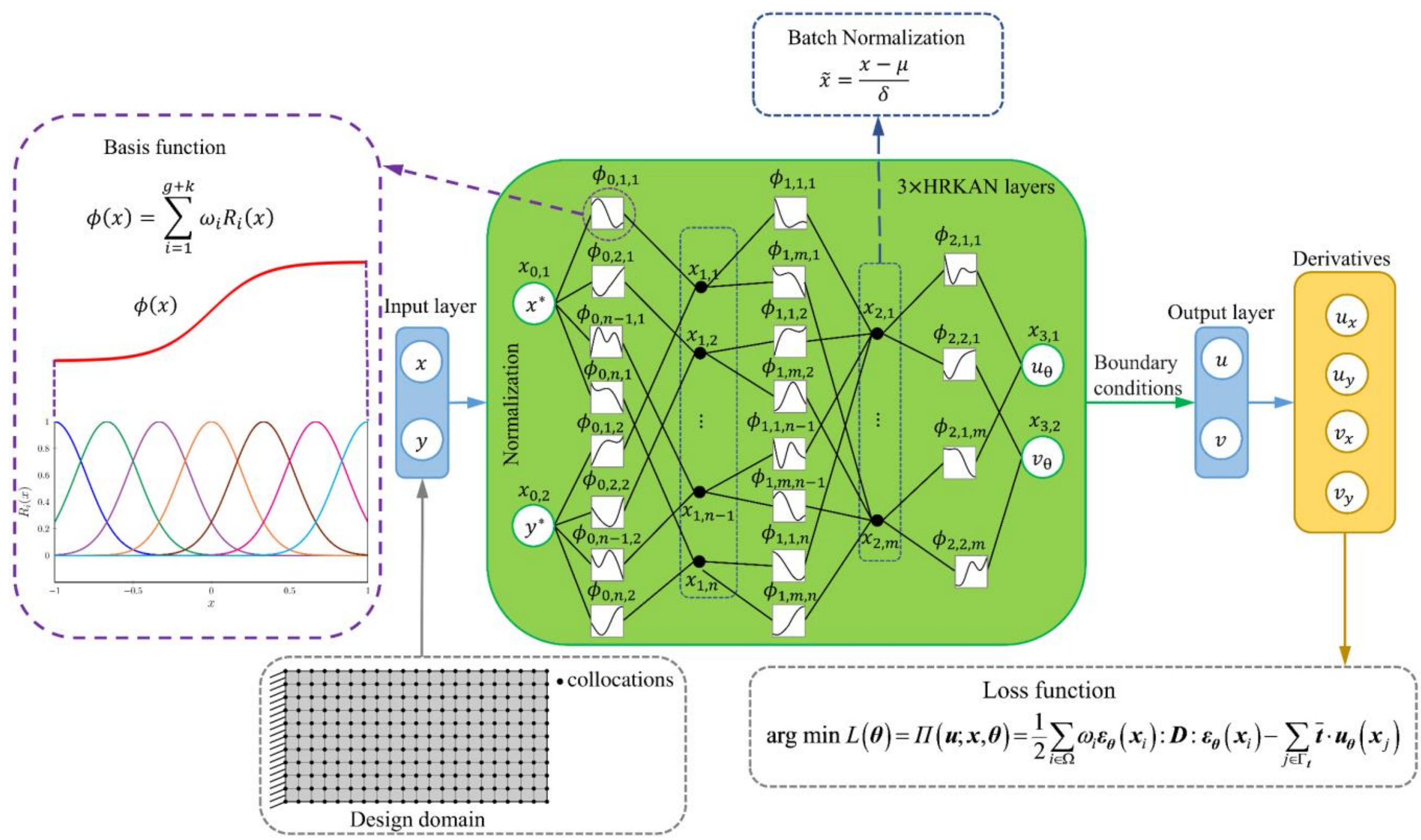

**Fig. 1** Schematic representation of the d-HRKAN. The network utilizes the coordinates ($x$, $y$) of sampled collocation points within the design domain as inputs and generates the displacement field ($u$, $v$) through a three-layer HRKAN architecture, which is trained by minimizing the total potential energy.

### *3.2 The architecture of s-HRKAN*

Topology optimization is fundamentally an iterative process of updating design variables to minimize the objective function. Building on this paradigm, transforming the objective function into a neural network loss function and minimizing it through the optimization of network parameters constitutes a commanding and viable strategy (Zhang et al. 2021). To reveals this approach, this study illustrates the implementation details of the s-HRKAN framework, using the 2D minimum compliance design problem as a representative example.

It is important to highlight that s-HRKAN architecture, illustrated in Fig. 2, bears resemblance to the sensitivity analysis network S-PINN utilized in CPINNTO (Jeong et al. 2023). However, a fundamental architectural distinction exists in the underlying neural network structure. In the proposed s-HRKAN framework, the iterative update of design variables is directly facilitated by updating the parameters of s-HRKAN. To clarify this procedure, consider the standard design problem of minimizing structural compliance subject to volume constraints as an illustrative example. The mathematical formulation for this problem is as follows:

$$\begin{aligned} &\min \; C(\boldsymbol{\theta}) = \boldsymbol{u}^{\mathrm{T}} \boldsymbol{K} \boldsymbol{u} \\ &\text{s.t.} \;\; \boldsymbol{K}(\boldsymbol{\theta}) \boldsymbol{u} = \boldsymbol{F} \\ &\qquad f_{\mathrm{v}}(\boldsymbol{\rho}) = \frac{\sum_{i \in \mathbb{N}_{\mathrm{e}}} \rho_i v_i}{V} = V_f \\ &\qquad 0 \le \rho_i \le 1, \;\; i \in \mathbb{N}_{\mathrm{e}}, \end{aligned} \tag{16}$$

where, $\boldsymbol{F}$ denotes the global force vector, and $\boldsymbol{K}$ is the global stiffness matrix. $V_f$ represents the prescribed volume constraint, while $\rho_i$ denotes the density of element $i$. Following the approach proposed by Chandrasekhar et al. (2021), the loss function of the s-HRKAN is formulated by incorporating the constraint into the objective function via the penalty function method:

$$\arg\min L_{\mathrm{obj}}(\boldsymbol{\theta}) = \mathcal{L}_C + \alpha \mathcal{L}_{V_f} = \frac{\boldsymbol{u}^{\mathrm{T}} \boldsymbol{K} \boldsymbol{u}}{C_0} + \alpha \left( \frac{f_{\mathrm{v}}(\boldsymbol{\rho})}{V_f} - 1 \right)^2, \tag{17}$$

where, $\alpha$ denotes the penalty coefficient, and $C_0$ denotes the initial structural compliance, which serves to scale the compliance gradient. By utilizing the backpropagation mechanism inherent in neural networks, this framework facilitates the updating of network parameters through automatic differentiation of the loss function gradient (Jeong et al. 2023).

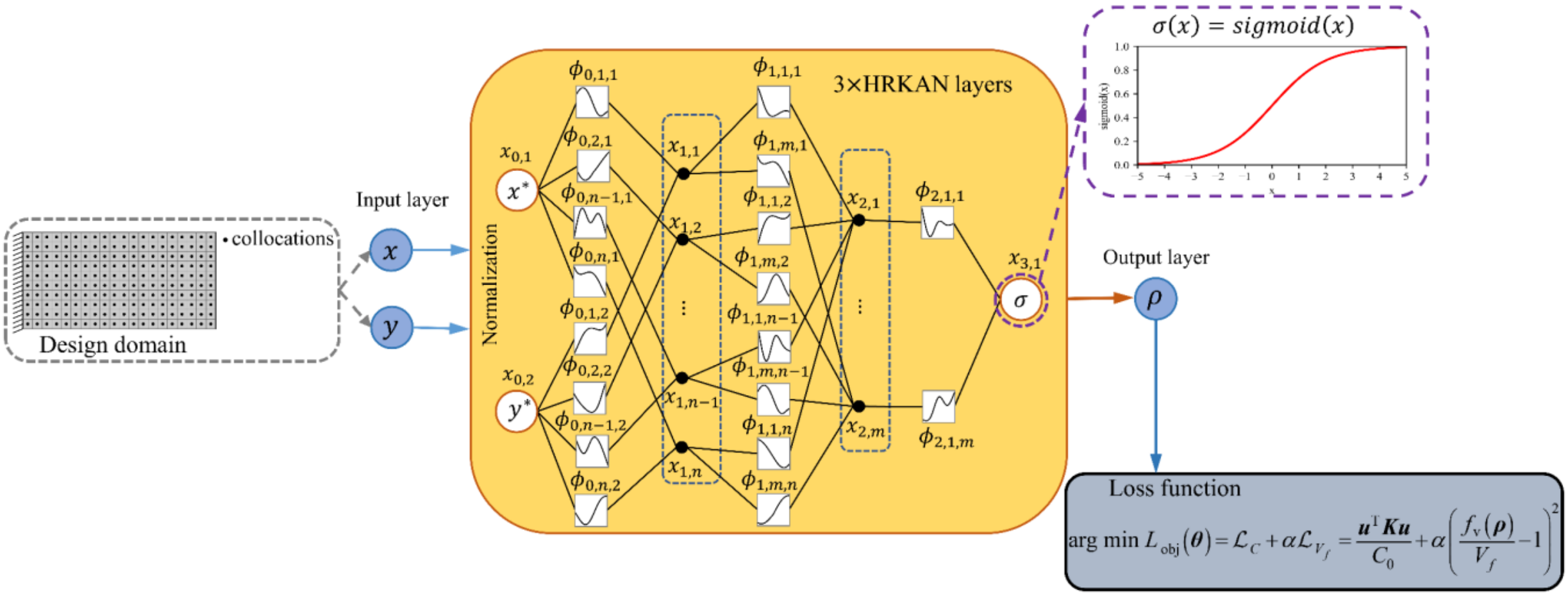

**Fig. 2** Schematic representation of the s-HRKAN architecture for topology optimization. The network receives the coordinates ($x$, $y$) of sampling points within the design domain as input and produces element-wise density $\rho$ as output through a three-layer HRKAN architecture. The loss function is formulated as a penalized objective function.

### *3.3 The implementation details of the DPIKAN-TO framework*

As depicted in Fig. 3, the DPIKAN-TO framework differentiates itself from traditional topology optimization methods by integrating d-HRKAN and s-HRKAN into the optimization process. The d-HRKAN is used to approximate the PDE-governed equilibrium-response field, while the s-HRKAN represents the sensitivity-driven topology-evolution field. Through density-dependent stiffness interpolation, displacement prediction, strain-energy evaluation, and sensitivity propagation, these two neural fields are coupled within a unified differentiable optimization framework. Therefore, DPIKAN-TO does not merely replace conventional finite element analysis (FEA) and sensitivity analysis procedures(Jeong et al. 2023), but reformulates topology optimization as a coupled neural-field evolution process.

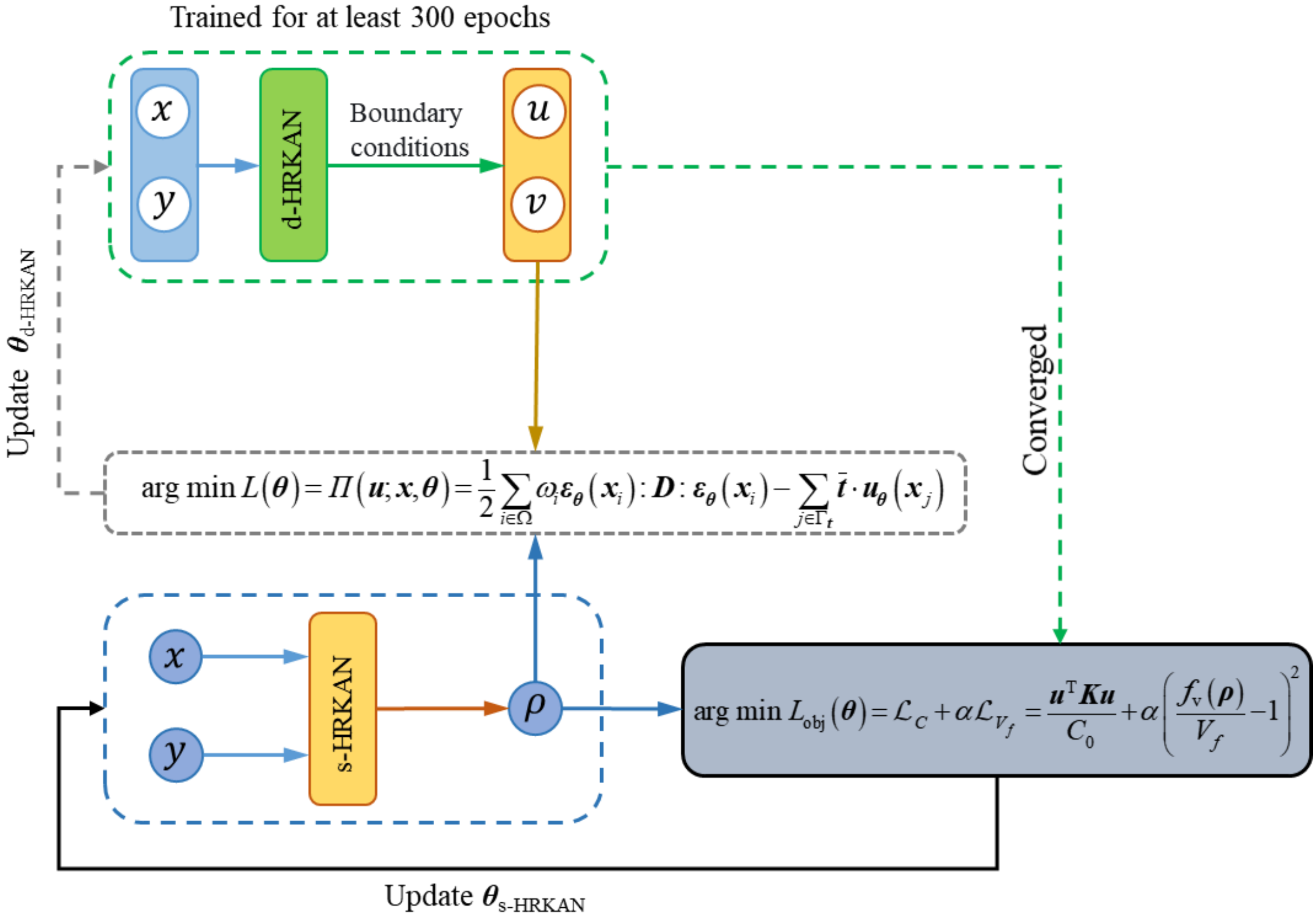

**Fig. 3** Schematic of the proposed DPIKAN-TO architecture for topology optimization. The s-HRKAN takes the spatial coordinates $(x, y)$ of sampling points as input and predicts the design variable $\boldsymbol{\rho}$ through forward propagation. The resulting density field $\boldsymbol{\rho}$ is then embedded into d-HRKAN, which is trained for at least 300 iterations to obtain an accurate displacement field corresponding to the current structural layout. Finally, the displacement field $(u, v)$ and the design variable $\boldsymbol{\rho}$ are used to compute the loss function of the s-HRKAN, and its parameters are updated via backpropagation, completing a single topology optimization iteration.

Based on the material interpolation model presented in Eq. (3), we compute Young's modulus $E(\rho_e)$ for the corresponding element, utilizing the density field $\boldsymbol{\rho}$ generated by s-HRKAN. Within the proposed DPIKAN-TO framework, upon achieving convergence, the resulting design adheres to the specified volume fraction and other imposed constraints while simultaneously minimizing the objective function. The convergence criterion employed in this study is as follows:

$$\frac{N_{\text{gray}}}{N} \le \tau , \tag{18}$$

where, $N_{gray}$ and $N$ represent the number of gray elements (with densities ranging from 0.05 to 0.95) and the total number of elements within the design domain. Additionally, $\tau$ denotes the specified convergence tolerance. To further clarify the role of neural representation in topology generation, Fig. 4 compares the density-field prediction mechanisms of MLP and KAN under the same coordinate-based input setting. Both models take spatial coordinates sampled from the design domain as inputs and predict the corresponding density field. The key difference lies in the activation mechanism: MLP employs fixed activation functions, whereas KAN uses localized learnable activation functions. This architectural difference enables KAN to provide a more flexible representation of sharp density transitions and localized structural features, which is particularly important for topology optimization problems involving solid–void interfaces and rapidly varying sensitivity fields.

| Model | Multi-Layer Perceptron | Kolmogorov-Arnold Network |
| --- | --- | --- |
| Density field prediction | | |
| Neural representation | | |
| Input coordinates (x, y) | • collocations | |

**Fig. 4** Schematic comparison of topology generation using MLP and KAN representations.

To enhance understanding of the optimization process, the overall workflow of the DPIKAN-TO framework is illustrated as follows:

---

**Algorithm:** Flowchart of the DPIKAN-TO framework for topology optimization

---

**Input :** $L_\Omega$, $W_\Omega$, $nelx$, $nely$, $\nu$, $\tau$, $E_{solid}$, $E_{\min}$, $V_f$, $penal$, $epoch_u$, $epoch_\rho$
**Result:** Optimized topology $\rho_{\text{opt}}$

**1:** Generate collocation points in $\Omega$ for training;
**2:** Set $\alpha_{\text{init}} = 10,\ \delta_\alpha = 0.05,\ \alpha_{\max} = 50,\ \text{epoch} = 0$;
**3:** Initialize models $\mathcal{N}_\rho(x, y; \theta_\rho)$ and $\mathcal{N}_u(x, y; \theta_u)$;
**4:** Initialize optimizers: $\text{optimizer}_\rho$, $\text{optimizer}_u$;
**5:** **repeat**
**6:** Reset gradients of $\text{optimizer}_\rho$;
**7:** Acquire density $\rho \leftarrow \mathcal{N}_\rho(x, y; \theta_\rho)$;
**8:** **repeat**
**9:** Reset gradients of $\text{optimizer}_u$;
**10:** $\mathcal{N}_u(x, y; \theta_u) \to (u, v)$; apply hard BCs;
**11:** Compute $\sigma, \epsilon$ via elastic tensor $C$ and shape function $N$;
**12:** Compute strain energy $U(\rho, u, v)$ via Gaussian integration;
**13:** Compute potential energy $\Pi(\rho, u, v) \to Loss_u$;
**14:** Update $\theta_u \leftarrow \arg\min Loss_u$;
**15:** **until** $\text{epoch} < \text{epoch}_u$;
**16:** Acquire converged displacement $(u, v)$ via $\mathcal{N}_u$; apply BCs;
**17:** Compute compliance $C(\rho, u, v)$;
**18:** Compute penalized objective $Loss_\rho \leftarrow \mathcal{L}_C(\rho, u, v) + \alpha \mathcal{L}_{V_f}$;
**19:** Update $\theta_\rho \leftarrow \arg\min Loss_\rho$;
**20:** Clip gradients of $\mathcal{N}_\rho$;
**21:** $\text{epoch} \leftarrow \text{epoch} + 1$;
**22:** **if** $error \leq \tau$ **then**
**23:** **break**;
**24:** **end**
**25:** **until** $\text{epoch} < \text{epoch}_\rho$;
**26:** Acquire final density $\rho_{\text{opt}}$ and plot result;

---

## 4. Numerical Examples

In this section, the proposed DPIKAN-TO framework is validated through a series of topology optimization problems governed by various types of PDEs, thereby demonstrating its efficiency and wide applicability. The framework is constructed by the PyTorch library, and the following benchmark problems are considered:

(1) Feasibility verification and computational efficiency analysis:

The performance of DPIKAN-TO is evaluated against the conventional topology optimization method on standard 2D topology optimization problems, focusing on structural outcomes and computational efficiency.

(2) Effects of mesh resolution and network parameters:

The influence of network hyperparameters and mesh resolutions on the final topology generated by DPIKAN-TO is systematically investigated.

(3) Topology optimization under complex design conditions:

① Topology optimization with stress constraints:

The proposed framework is utilized in topology optimization problems that incorporate stress constraints, thereby demonstrating its efficacy in addressing intricate design requirements.

② Topology optimization with design-dependent fluid pressure loads:

To evaluate its applicability, the framework is implemented in cases that involve fluid pressure loads dependent on design parameters.

③ Three-Dimensional topology optimization problems:

To evaluate its applicability in 3D problems, the proposed framework is implemented on a standard example of topology optimization in three dimensions.

### *4.1 Feasibility verification and computational efficiency analysis*

To assess the feasibility and computational efficiency of the proposed DPIKAN-TO framework, this section presents a comparison between standard 2D topology optimization results generated by DPIKAN-TO and those obtained using the SIMP method. The reference SIMP cases are implemented based on a 110-line MATLAB code that incorporates Heaviside projection, as originally proposed by Andreassen et al. (2011).

A deep beam with non-dimensionalized dimensions of 2×1 is used as the design domain. It is discretized into 80×40 elements, yielding an element length of $L_e = 0.025$. The material properties are specified as follows: Young's modulus $E$ =1, Poisson's ratio $\nu$ = 0.3, and $E_{\min} = 1\times10^{-6}$. The concentrated load with a magnitude of $F$ = 1, and distributed load of $f$ = 1 imposed on the deep beam. The topology optimization parameters used in the work by Andreassen et al. (2011) are set as follows: the density penalization factor is $p$=3, the density filter radius is $r_{\min}$ = 3, and the convergence tolerance (as defined in Eq. (18)) is $\tau$ = 0.05.

The hyperparameter configurations for the networks are detailed as follows. The d-HRKAN utilizes a network architecture of [2, 32, 32, 2], with the basis function configuration parameters set to grid points $G$ = 5, basis function order $k$ = 4, and $\beta$ = 3. Meanwhile, the s-HRKAN employs an architecture of [2, 40, 40, 1], with its basis function configuration specified as $G$ = 6, $k$ = 4, $\beta$ = 3. For both networks, the support intervals of the basis function grid are initialized within the range of [-1,1]. The maximum number of training iterations for the d-HRKAN is set to 1000, with an early stopping strategy implemented after 300 iterations to enhance convergence efficiency. The initial learning rate for both networks is established at $2\times10^{-3}$. For s-HRKAN, the penalty coefficient associated with the constraint term is initialized as $\alpha$=10, with an increment size of 0.05 and a maximum limit set at $\alpha_{max}$=50. It should be emphasized that an overly small penalty factor allows noticeable constraint violations to persist in the final design, whereas an excessively large initial α makes the optimization overly stiff and prevents intermediate densities from being reduced effectively. The adopted continuation strategy provides a compromise by allowing more flexible exploration in the early stages and stricter constraint enforcement near convergence.

The optimization results presented in Fig. 5 demonstrate that the topology generated by DPIKAN-TO exhibits distinct and smooth boundaries. In contrast, the SIMP designs yield blurred structural boundaries due to the presence of intermediate-density regions, which arise from an overestimation of strain energy (Guest et al. 2004). To facilitate a more equitable and intuitive comparison, the Heaviside projection filter is applied to the SIMP designs. This approach effectively mitigates compliance discrepancies caused by gray elements within the continuous design representation (Sigmund 2022). It is important to highlight that, even in the absence of filtering the density field, the designs produced by DPIKAN-TO do not display a checkerboard effect (Chandrasekhar and Suresh 2021). Since the basis functions in HRKAN directly enhance the network's capacity to represent high-frequency features, DPIKAN-TO framework can partially mitigate the oversimplified structural designs that are frequently observed in PINN-based topology optimization methods. This oversimplification is often attributed to the spectral bias inherent in MLP architectures (Jeong et al. 2023;Sanu et al. 2024). In contrast, the HRKAN used in DPIKAN-TO exhibits a reduced spectral bias and is better able to capture high-frequency structural features(Wang et al. 2024). As a result, especially at low volume fractions,

DPIKAN-TO produces more slender load-bearing members that are comparable to those obtained by the SIMP method. Besides, although the analysis mesh is 80 × 40, the smooth boundaries shown in Fig. 5 are obtained without any additional training. This is accomplished by evaluating the pre-trained model on a significantly finer set of sampling points and subsequently plotting the resulting iso-contour of the continuous density field, which provides a high-resolution representation.

Notably, as illustrated in Table 1, the compliance difference between structures generated by DPIKAN-TO and those produced using the SIMP method in standard 2D topology optimization benchmarks do not exceed 1.48%. This finding indicates a close agreement in their final performance. Furthermore, the DPIKAN-TO framework consistently necessitates a significantly lower number of iterations to achieve convergence when compared to the SIMP method utilizing Heaviside filtering.

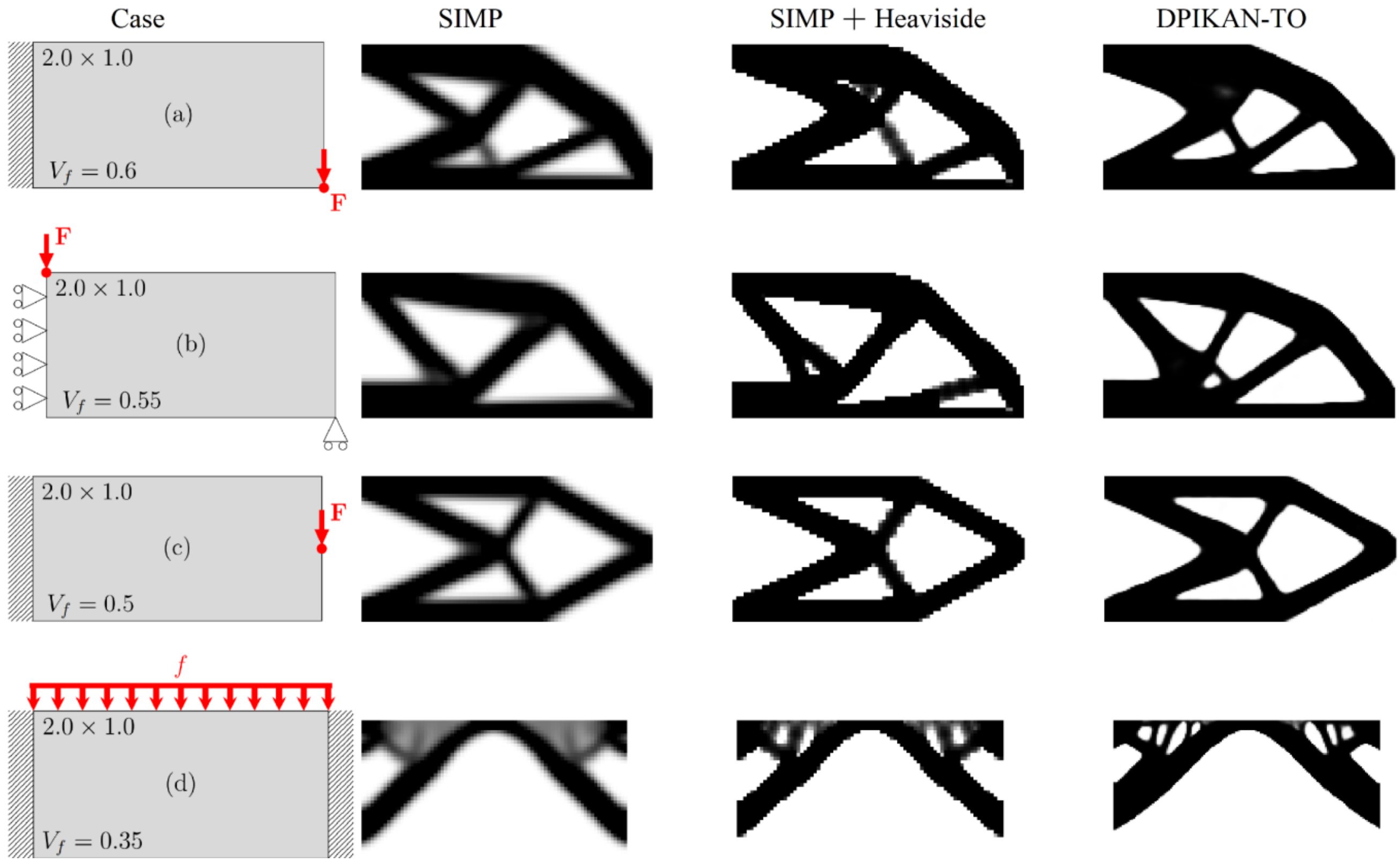

**Fig. 5** Validation of DPIKAN-TO against standard minimum compliance problems with different boundary and loading conditions, using SIMP as reference.

`

**Table 1**

Comparison of Numerical Results obtained by DPIKAN-TO and SIMP.

| Case | Compliance | | | | Iterations | | |
|---|---|---|---|---|---|---|---|
| | SIMP | SIMP + Heaviside | DPIKAN-TO | % diff | SIMP | SIMP + Heaviside | DPIKAN-TO |
| (a) | 64.25 | 60.34 | 60.64 | 0.49 | 194 | 522 | 120 |
| (b) | 77.34 | 71.40 | 71.72 | 0.45 | 57 | 801 | 129 |
| (c) | 69.40 | 62.38 | 62.50 | 0.19 | 44 | 430 | 97 |
| (d) | 2.46 | 2.04 | 2.07 | 1.48 | 32 | 435 | 201 |

As illustrated in the convergence history of the cantilever beam with a tip-centered point load (Fig. 6), the substantial initial penalty imposed on the volume constraint renders the network more sensitive to violations of this constraint than to compliance considerations. Consequently, during optimization, the volume fraction converges to its constraint value earlier than does compliance. As illustrated in the computational efficiency comparison presented in Table 2, although DPIKAN-TO framework necessitates fewer iterations to achieve convergence compared to the SIMP method utilizing Heaviside filtering, its overall computational time remains significantly higher—approximately 20 to 43 times greater in standard 2D topology optimization problems. This discrepancy is primarily attributed to the requirement for d-HRKAN to undergo training for a minimum of 300 iterations following each design update in order to accurately predict the displacement field. This is due to the fact that, in linear elastic structural analysis, neural networks necessitate numerous iterations to solve the governing PDEs, whereas the finite element method can achieve a solution in a single analysis step.

Under the same 2D design problems and convergence criteria, a comparative analysis of computational efficiency is performed between DPIKAN-TO and CPINNTO framework. According to the benchmark data reported by Jeong et al. (2023) (refer to Appendix A.1), the MLP-based CPINNTO framework necessitates a total computational time that is approximately 34 to 102 times greater than that required by the SIMP method with Heaviside projection filter. In contrast, the proposed DPIKAN-TO framework exhibits significantly enhanced efficiency compared to CPINNTO when applied to identical problems.

**Table 2**

Comparison of computational efficiency between DPIKAN-TO and SIMP.

| Case | Average CPU time (seconds) | | | Total CPU time (seconds) | | | |
|---|---|---|---|---|---|---|---|
| | SIMP | SIMP+Heaviside | DPIKAN-TO | SIMP | SIMP+Heaviside | DPIKAN-TO | $t_{\mathrm{DPIKANTO}}/t_{\mathrm{SIMP+Heaviside}}$ |
| (a) | 0.035 | 0.038 | 4.412 | 6.890 | 20.029 | 529.474 | 26.435 |
| (b) | 0.038 | 0.039 | 4.822 | 2.171 | 31.261 | 622.059 | 19.899 |
| (c) | 0.044 | 0.039 | 4.487 | 1.979 | 16.588 | 435.210 | 26.236 |
| (d) | 0.046 | 0.046 | 4.312 | 1.472 | 20.182 | 866.670 | 42.943 |

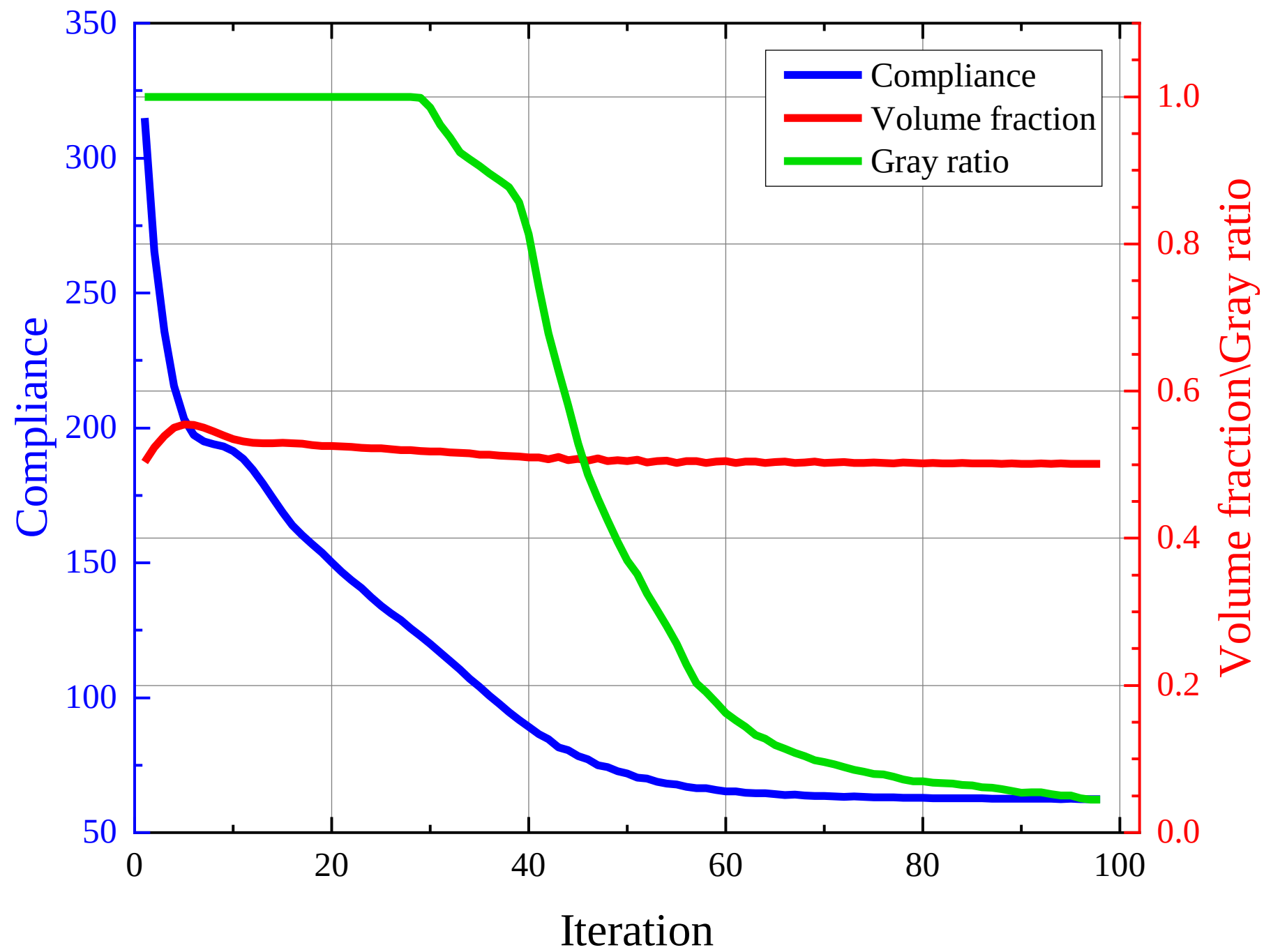

**Fig. 6** DPIKAN-TO convergence history for the structure in Fig. 5(c).

### *4.2 Effects of mesh resolution and network structure*

This section investigates the influence of network structure and mesh resolution on the topology designs generated by DPIKAN-TO.

#### *4.2.1 Effect of network structure on topology design*

Due to the highly nonlinear nature of neural network parameters (Chandrasekhar and Suresh 2021), evaluating the impact of network architecture presents a significant challenge. To address this issue, the current subsection explores the influence of various network structure configurations on both topology and convergence behavior in the cantilever beam with a tip-centered point load (Section 4.1), by exclusively varying the architecture of s-HRKAN. As demonstrated in (Z. Liu et al. 2024), a two-layer composition of KAN is theoretically sufficient for accurately representing any continuous multivariate function. However, in practical applications, deeper KAN architectures are frequently utilized to enhance training stability. Given the propensity of KAN to overfit during training (Sohail 2024), a network depth of two to three layers is generally adequate for producing effective topological designs for benchmark problems.

As illustrated by the topological configurations and convergence results of DPIKAN-TO across various s-HRKAN architectures, as presented in Fig. 7 and Table 3, increasing the network depth from two to three layers markedly reduces the number of iterations and computational time required by DPIKAN-TO (Fig. 7 (a) and (b)). However, as illustrated in Table 3, an increase in network depth leads to a temporary rise in structural compliance. This phenomenon can primarily be attributed to the reduction of network parameters and the expedited convergence process, which collectively cause the network to produce topology layouts exhibiting slight asymmetry. By broadening the network layers (i.e., increasing the number of parameters within each layer), the convergence process can be further accelerated, while the asymmetry in the layouts is mitigated, thereby reducing the compliance error in the final designs (Fig. 7 (b) – (d)).

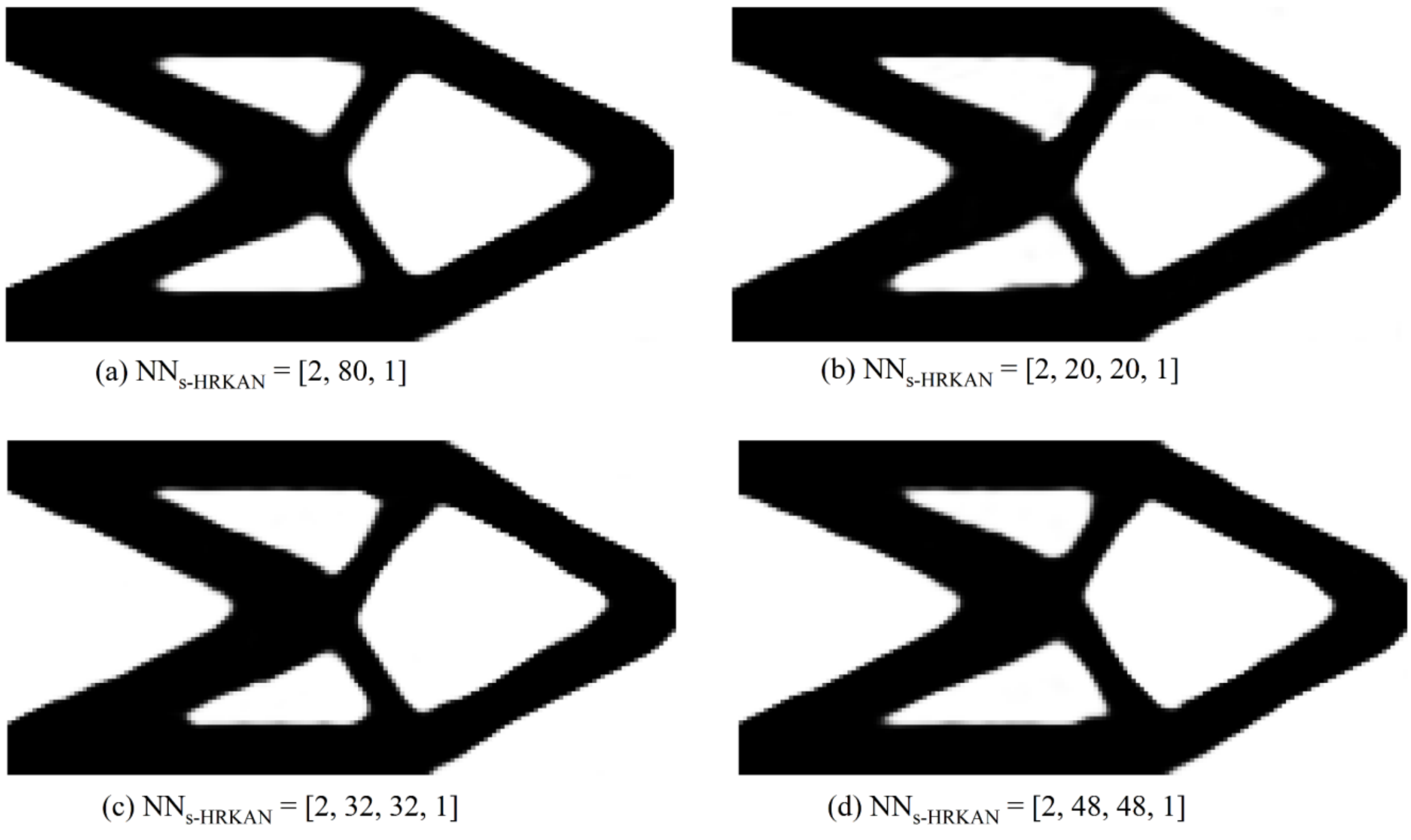

Fig. 7 Optimal topologies obtained by DPIKAN-TO under different s-HRKAN architectures.

**Table 3**

Convergence results of DPIKAN-TO under different s-HRKAN architectures.

| Case | Neural network | Iterations | Objective value | Total CPU time (s) |
|---|---|---|---|---|
| (a) | [2, 80, 1] | 499 | 62.60 | 2239.013 |
| (b) | [2, 20, 20, 1] | 144 | 62.80 | 646.129 |
| (c) | [2, 32, 32, 1] | 112 | 62.56 | 502.545 |
| (d) | [2, 48, 48, 1] | 90 | 62.54 | 403.83 |

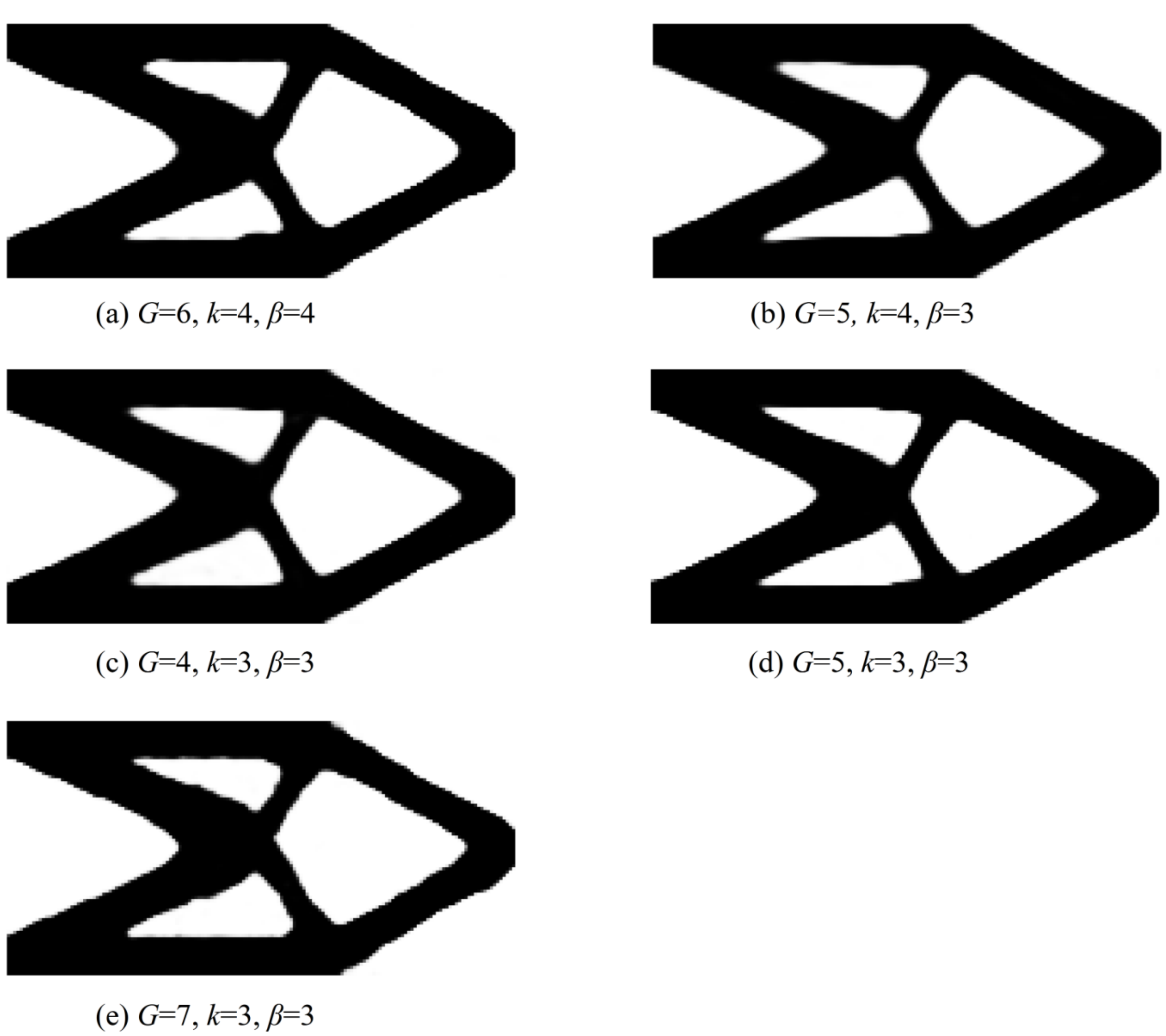

(a) $G$=6, $k$=4, $\beta$=4

(b) $G$=5, $k$=4, $\beta$=3

(c) $G$=4, $k$=3, $\beta$=3

(d) $G$=5, $k$=3, $\beta$=3

(e) $G$=7, $k$=3, $\beta$=3

Fig. 8 Optimal topology designs generated by DPIKAN-TO under different combinations of hyperparameters ($G$, $k$, $\beta$), with the s-HRKAN structure fixed as $NN_{s\text{-}HRKAN}$ = [2, 40, 40, 1].

Moreover, as illustrated in Table 4, various combinations of hyperparameters ($G$, $k$, $\beta$) (see Eq. (6) and Eq. (7)) also affect the convergence performance of DPIKAN-TO. An increase in grid resolution $G$ and basis function order $k$ leads to improvements in both structural resolution and convergence speed. Conversely, the parameter $\beta$ does not enhance the quality of the resulting topology or its structural stiffness; rather, it tends to diminish the convergence rate. It is important to note that while the impact of hyperparameters on convergence is not as pronounced as that of the network architecture, appropriate parameter settings can still mitigate design asymmetry while maintaining convergence speed (refer to Fig. 8 (b) and (d)). Consequently, in addition to considering the network structure, the selection of hyperparameters ($G$, $k$, $\beta$) is also crucial to the performance of topology optimization.

**Table 4**

Convergence results of DPIKAN-TO under different combinations of hyperparameters ($G$, $k$, $\beta$).

| Case | Hyperparameter ($G$, $k$, $\beta$) | Iterations | Objective value | Total CPU time (s) |
|---|---|---|---|---|
| (a) | (6, 4, 4) | 106 | 62.41 | 475.623 |
| (b) | (5, 4, 3) | 92 | 62.44 | 412.805 |
| (c) | (4, 3, 3) | 129 | 62.64 | 578.823 |
| (d) | (5, 3, 3) | 106 | 62.66 | 475.625 |
| (e) | (7, 3, 3) | 135 | 62.94 | 605.754 |

#### *4.2.2 Effect of finite element mesh resolution on topology design*

Since the total potential energy in d-HRKAN is calculated using Gaussian quadrature, it requires the discretization of the design domain within the loss function (Li et al. 2021). Consequently, the d-HRKAN also demonstrates the characteristic of "mesh dependency". Therefore, it is crucial to examine the impact of varying finite element mesh resolutions on the performance of DPIKAN-TO.

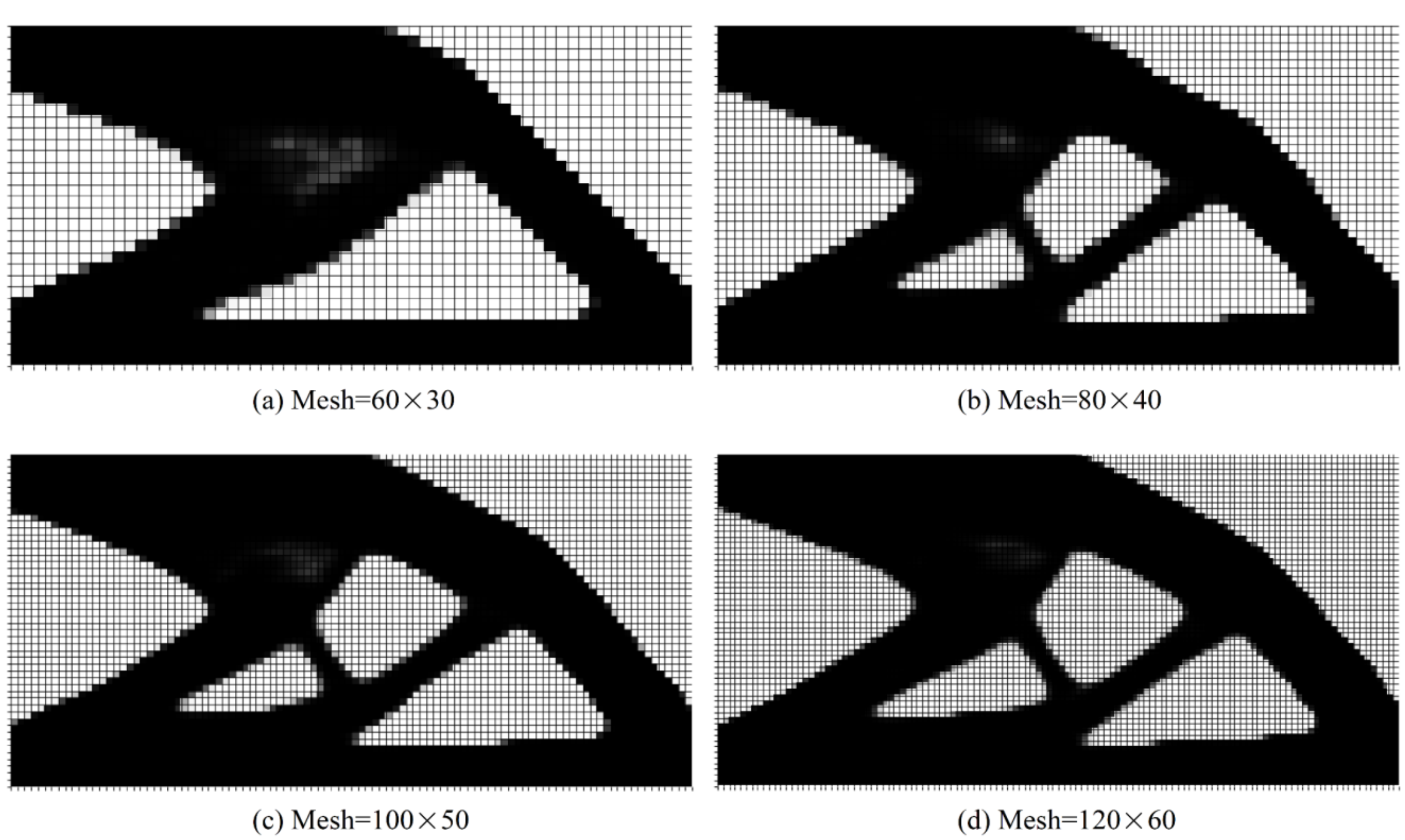

**Fig. 9** Optimal topologies designed by DPIKAN-TO under different mesh resolutions with s-HRKAN and d-HRKAN architectures fixed as $NN_{s\text{-}HRKAN}$ = [2, 40, 40, 1] and $NN_{d\text{-}HRKAN}$ = [2, 32, 32, 2], respectively.

As illustrated in Fig. 9, the topologies of the cantilever beam example with a tip load (Section 4.1), designed by DPIKAN-TO under varying mesh resolutions, demonstrate a progressive decrease in compliance and increasingly refined structural characteristics as the resolution enhances. This phenomenon can be attributed to the mesh-dependence inherent in the framework. As shown in Table 5 and Fig. 10, the convergence results of DPIKAN-TO under varying mesh resolutions reveal that, as the mesh resolution increases, the number of sampling points that d-HRKAN must process in each iteration significantly escalates, resulting in a considerable rise in computational cost per iteration. Concurrently, the number of iterations required for convergence tends to increase with finer resolutions, thereby diminishing the overall convergence speed and substantially elevating the computational expense of DPIKAN-TO.

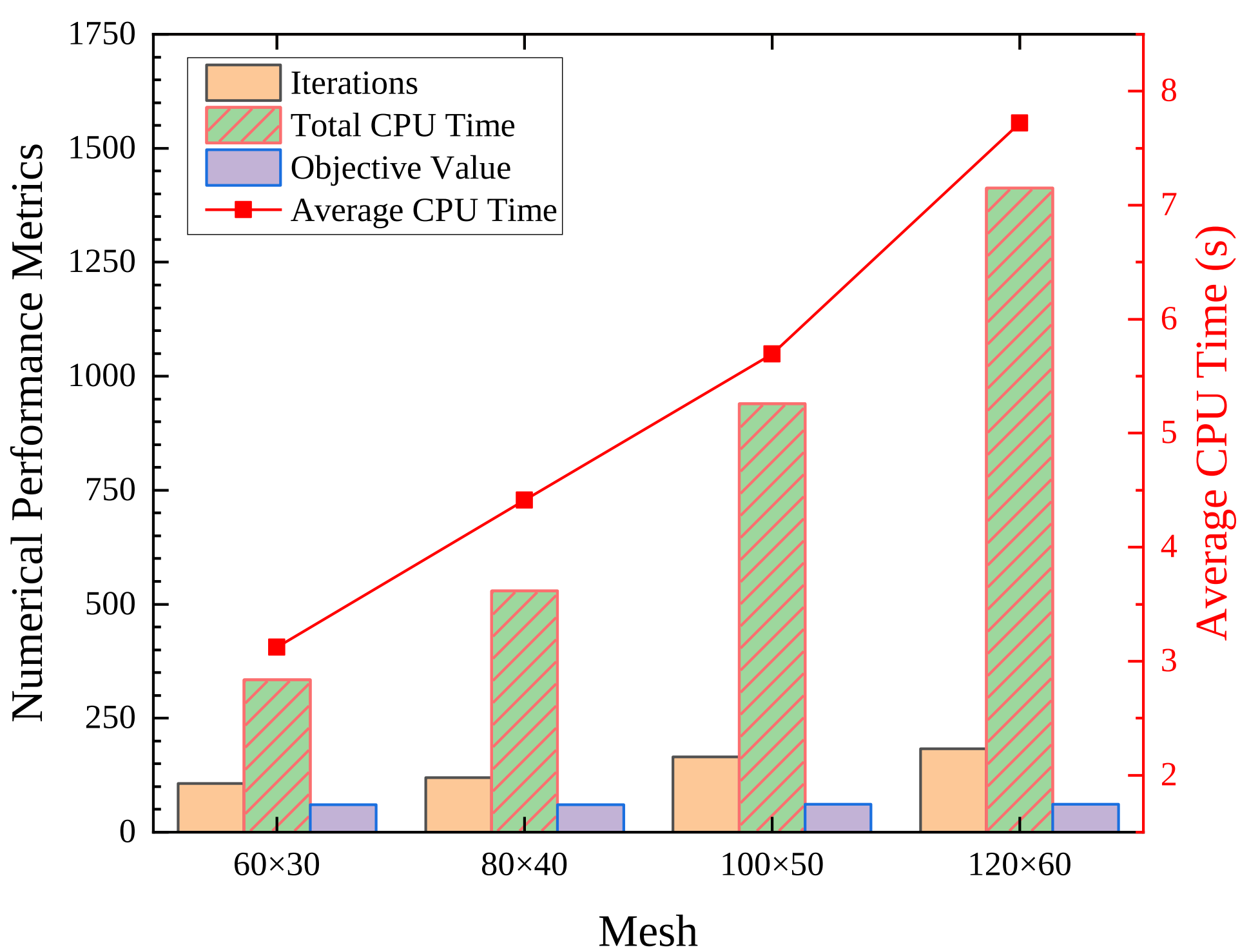

**Fig. 10** Comparison of convergence behavior and compliance values of DPIKAN-TO with varying mesh resolutions.

**Table 5**

Convergence results of topology designs under different mesh resolutions.

| Case | Iterations | Objective value | Total CPU time (s) | Average CPU time (s) |
|---|---|---|---|---|
| (a) | 107 | 60.69 | 333.923 | 3.121 |
| (b) | 120 | 60.64 | 529.474 | 4.412 |
| (c) | 165 | 60.88 | 939.645 | 5.695 |
| (d) | 183 | 61.17 | 1412.493 | 7.719 |

### *4.3 Topology optimization with stress constraints*

This section explores the applicability and performance of DPIKAN-TO in addressing strongly nonlinear and non-convex optimization problems, by solving a minimum compliance problem subject to both volume and maximum von Mises stress constraints.

#### *4.3.1 Implementation details of DPIKAN-TO for stress-constrained topology optimization*

The mathematical formulation for minimizing structural compliance while adhering to both a volume constraint and a maximum von Mises stress constraint can be articulated as follows:

$$
\begin{aligned}
&\min \;\; C = \boldsymbol{u}^{\mathrm{T}} \boldsymbol{K} \boldsymbol{u} \\
&\text{s.t.} \;\; \boldsymbol{Ku} = \boldsymbol{F} \\
&\qquad f_{\mathrm{s}}(\boldsymbol{\rho}) = \max(\boldsymbol{\sigma}_{\mathrm{vm}}) \le \sigma_{\mathrm{vm}}^{*} \\
&\qquad f_{\mathrm{v}}(\boldsymbol{\rho}) = \frac{\sum_{i \in \mathbb{N}_{\mathrm{e}}} \rho_i v_i}{V} = V_f \\
&\qquad 0 \le \rho_i \le 1, \;\; i \in \mathbb{N}_{\mathrm{e}},
\end{aligned}
\tag{19}
$$

where, $\boldsymbol{\sigma}_{\mathrm{vm}}$ and $\sigma_{\mathrm{vm}}^{*}$ denote the elemental von Mises stress and the stress threshold that incorporates safety redundancy. To mitigate stress singularities in void elements, a stress penalization technique is employed to relax the stress constraints (Holmberg et al. 2013):

$$
\sigma_{\mathrm{vm},i} = x_i^{q} \sigma_{\mathrm{vm},i} \, , \tag{20}
$$

where, $\sigma_{\mathrm{vm},i}$ represents the relaxed von Mises stress in the $i$-th element, while $q$ denotes a non-negative penalization factor. Furthermore, due to the non-differentiability of the maximum stress as expressed by a max-function, the N-norm stress aggregation method proposed by Le et al. (2010) is commonly utilized to approximate the maximum stress in a smooth and differentiable manner:

$$
\sigma_{\mathrm{PN}} = \left( \sum_{i=1}^{n} \sigma_{\mathrm{vm},i} \right)^{\frac{1}{N}} . \tag{21}
$$

As the integer parameter $N$ increases from 1 to infinity, the $N$-norm stress measured gradually transitions from reflecting average stress to closely approximating maximum stress. To address issues related to stress singularities while maintaining compliance minimization, we introduce the distance function between max($\boldsymbol{\sigma}_{\mathrm{vm}}$) and $\sigma_{\mathrm{vm}}^{*}$, denoted as the stress constraint expression $f_{\mathrm{s}}(\boldsymbol{\rho})$. This distance measure is incorporated into the loss function of s-HRKAN through a fixed penalization factor $\alpha_{\mathrm{stress}}$. Accordingly, the resulting s-HRKAN loss function can be expressed as:

$$
\arg\min \; L_{\mathrm{obj}}(\boldsymbol{\theta}) = \frac{\boldsymbol{u}^{\mathrm{T}} \boldsymbol{K} \boldsymbol{u}}{C_0} + \alpha \left( \frac{f_{\mathrm{v}}(\boldsymbol{\rho})}{V_f} - 1 \right)^2 + \alpha_{\mathrm{stress}} \sqrt{\left( \sigma_{\mathrm{PN}} - \sigma_{\mathrm{vm}}^{*} \right)^2} \, . \tag{22}
$$

Consequently, the final topology not only fulfills the objective of compliance minimization but also effectively mitigates the occurrence of stress singularities.

#### *4.3.2 Topology Optimization Benchmark with Stress Constraints*

In this section, an L-shaped bracket example is utilized to assess the performance of DPIKAN-TO in tackling strongly nonlinear and non-convex topology optimization problems. To validate the feasibility of the DPIKAN-TO designs for this specific issue, we modified the 146-line stress sensitivity analysis code originally proposed by Deng et al. (2022). Specifically, we replaced the original objective function with compliance minimization and incorporated a global von Mises stress constraint, while maintaining all other simulation parameters unchanged. The modified code is subsequently employed to generate comparative designs against those produced by DPIKAN-TO. As illustrated in Fig. 11, the design domain of the L-shaped bracket is discretized into elements with a side length of $L_e$ = 0.05. The volume fraction constraint is established at $V_f$ = 0.4. The uniformly distributed force of magnitude $f$ = 4 is applied to eight sampling points along the designated boundary segment. All material properties and the parameters of topological optimization are consistent with those detailed in Section 4.1. In the parameter configuration of the stress constraint term, the penalty factor for stress loss $\alpha_{stress}$ is set to 0.2, the size of the norm is 10, and the threshold for stress constraints is defined as 35.

In stress-constrained topology optimization problems, the convergence criterion for DPIKAN-TO must be appropriately revised. The optimization process should only be terminated when both the original convergence conditions are met and the maximum von Mises stress of the structure is below the specified stress threshold. It is important to highlight that no modifications are necessary for the network architecture of the s-HRKAN when tuning hyperparameters. However, to improve the accuracy of displacement gradients in proximity to stress-concentration corners, the d-HRKAN is designed via $NN_{d\text{-}HRKAN}$ = [2, 40, 40, 2], while maintaining the original hyperparameter configuration. Furthermore, due to the increased mesh resolution, the early stopping iteration limit has been doubled to meet the heightened computational demands.

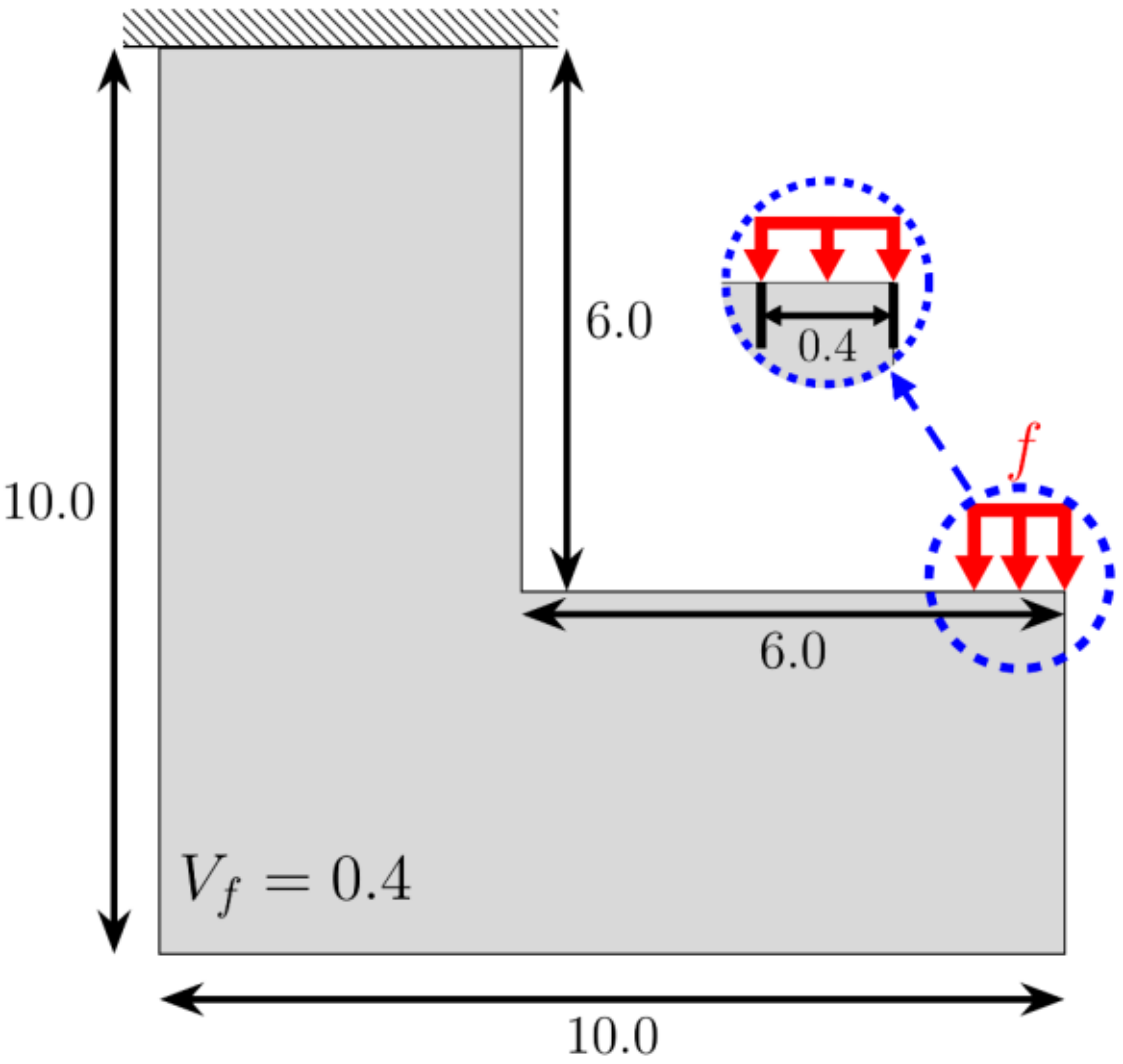

**Fig. 11** Design domain of the L-shaped beam used to test stress-constrained topology optimization. A distributed load of magnitude 4 is uniformly applied at eight sampling points along the upper-right boundary.

By comparing the topology designs derived from stress-constrained and standard compliance minimization approaches, as illustrated in Fig. 12, the final topologies resulting from the standard compliance minimization design exhibit significant stress concentration in the corner regions. This phenomenon, attributed to geometric singularities, often compromises the structural durability of high-stiffness designs despite their optimal compliance performance. It is noteworthy that, as illustrated in Fig. 12, the stress field responses of the final topologies generated by both DPIKAN-TO and SIMP exhibit analogous structural modifications upon the application of the stress constraint. Specifically, the sharp corners present in the original designs are significantly smoothed effectively mitigating stress concentrations in those regions (Chen et al. 2021).

To further assess the influence of stress constraints on structural performance, the convergence results summarized in Table 6 indicate that the maximum von Mises stress at the corner region decreases from 51.9 to 25.8 for the DPIKAN-TO design and from 50.3 to 26.4 for the SIMP design, respectively. Under structural stress safety constraints, the compliance of stress-constrained topology optimization designs improved by about 3.3% and 8.5% compared to standard minimum compliance designs, while maintaining good structural stiffness performance. These findings further validate the feasibility and design efficiency of the proposed framework in stress-constrained topology optimization problems. They also highlight its potential for extension to strongly nonlinear systems and topologies that involve complex material behaviors. It is important to note that the compliance of designs with stress constraints is significantly higher than that of standard compliance minimization designs, as reducing peak stress often results in a compromise on structural stiffness (Yang and Chen 1996).

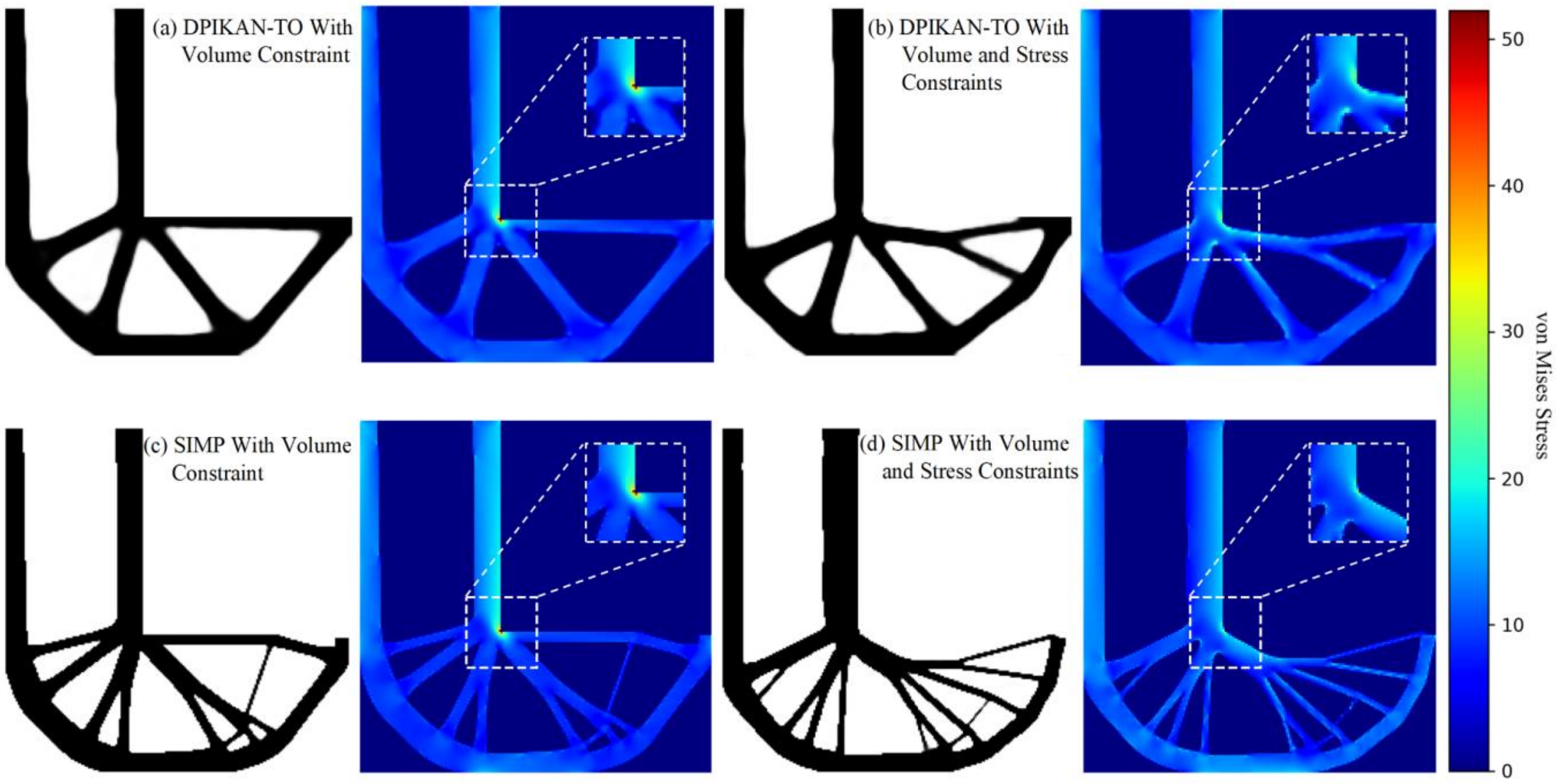

Fig. 12 Comparison of the final topologies and von Mises stress responses obtained by DPIKAN-TO and SIMP under different constraint settings.

**Table 6**

Convergence results of topology designs shown in Fig. 12.

| Case | Method | Iterations | Objective Value | Maximum von Mises Stress |
|---|---|---|---|---|
| (a) | DPIKAN-TO | 139 | 3157.30 | 51.9 |
| (b) | DPIKAN-TO | 333 | 3261.83 | 25.8 |
| (c) | SIMP | 141 | 3047.21 | 50.3 |
| (d) | SIMP | 198 | 3306.64 | 26.4 |

By analyzing the convergence history of the stress-constrained compliance minimization problem illustrated in Fig. 13, it is evident that the inclusion of the stress penalty term significantly influences the initial evolution of the structure towards satisfying the $N$-norm stress constraint. As a result, during the early stages of optimization, the volume fraction displays a non-monotonic trend. Concurrently, both structural compliance and peak von Mises stress consistently decrease throughout the iterations, indicating a progressive enhancement in stiffness while ensuring adherence to stress safety requirements. In the later iterations, both volume fraction and $N$-norm stress converge stably, satisfying the prescribed constraints and yielding a well-optimized topology. It is noteworthy that reducing the stress penalty factor or s-HRKAN learning rate can improve stability but slows convergence, requiring a balance between efficiency and constraint satisfaction.

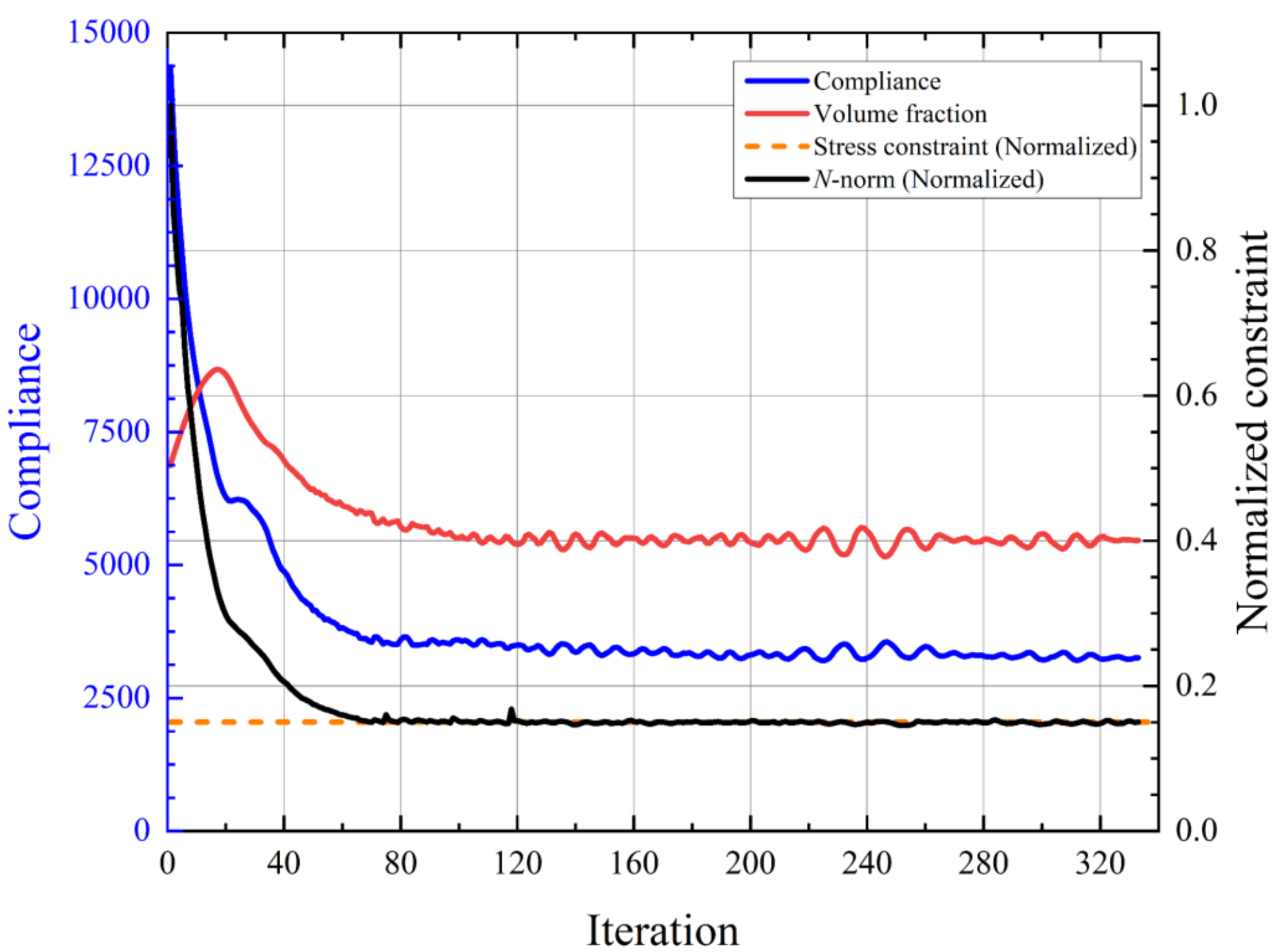

**Fig. 13** Convergence history of DPIKAN-TO for the topology optimization problem with stress constraints. The stress constraint values are normalized by their initial value.

## *4.4 Topology optimization with design-dependent fluid pressure loads*

### *4.4.1 Construction of pressure load field*

During the topology optimization process of fluid-structure interaction systems, the solid and void states of the elements undergo continuous changes with each iteration. This dynamic alteration results in variations in both the application area and direction of the pressure load. To prevent the need for repeated boundary tracking in each iteration, a continuous pressure field is established across the entire design domain and subsequently transformed into equivalent nodal structural loads, following the methodology proposed by Kumar et al. (2020). The fluid pressure loading is modeled according to Darcy's law, from which the volumetric flow rate of the porous medium can be derived as follows $\boldsymbol{q}$:

$$\boldsymbol{q} = -\frac{\kappa}{\mu}\nabla p = -K(\boldsymbol{\rho})\nabla p, \tag{23}$$

where, $\nabla p$ represents the pressure gradient, while $\kappa$ and $\mu$ denote the permeability of the porous medium and the fluid viscosity, respectively. The term $K(\rho_e)$ signifies the flow conductivity of element e. Besides, as depicted in Fig. 14(a), a residual pressure gradient is evident in the solid regions of the domain, despite an absence of actual flow. To address the issue of residual gradient, a drainage term $\boldsymbol{Q}_{\text{drain}}$ is incorporated into Darcy's law. This modification facilitates the generation of the anticipated pressure field distribution, as illustrated in Fig. 14(b). Consequently, the revised equilibrium equation can be articulated as follows (Kumar et al. 2020):

$$\nabla \cdot \boldsymbol{q} - \boldsymbol{Q}_{\text{drain}} = \nabla \cdot \left(K(\boldsymbol{\rho})\nabla p\right) + \boldsymbol{Q}_{\text{drain}} = 0, \tag{24}$$

where, $p$ denotes the continuous pressure field. The drainage term is defined as $\boldsymbol{Q}_{\text{drain}} = -D(\boldsymbol{\rho})(p-p_{\text{out}})$, where $D(\rho_e)$ denotes the drainage coefficient of element e.

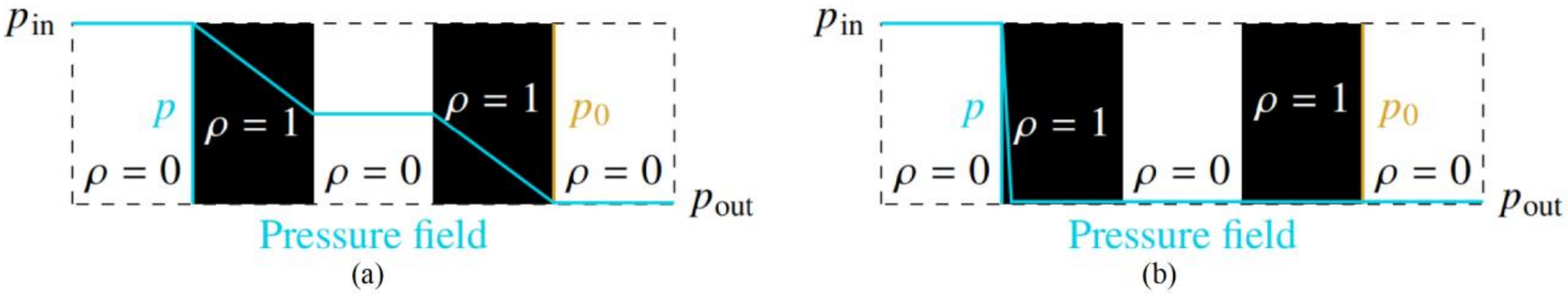

**Fig. 14** Comparison of pressure fields in solid regions with and without drainage term (Kumar 2023): (a) Pressure field without drainage; (b) Pressure field with drainage correction. Variables $p_{\text{in}}$ and $p_{\text{out}}$ represent the inlet and outlet pressures, respectively.

According to Kumar et al. (2020), the application of finite element discretization to Eq. (24), combined with the Galerkin approach, results in the following expression:

$$\int_{\Omega_e}\left(K\boldsymbol{B}_{\boldsymbol{P}}^{\text{T}}\boldsymbol{B}_{\boldsymbol{P}} + D\boldsymbol{N}_{\boldsymbol{P}}^{\text{T}}\boldsymbol{N}_{\boldsymbol{P}}\right)\text{d}V\boldsymbol{P}_{\text{e}} = \int_{\Omega_e}\left(D\boldsymbol{N}_{\boldsymbol{P}}^{\text{T}}p_{\text{out}}\right)\text{d}V - \int_{\Gamma_e}\left(\boldsymbol{N}_{\boldsymbol{P}}^{\text{T}}\boldsymbol{q}_{\Gamma}\cdot\boldsymbol{n}_{\text{e}}\right)\text{d}V, \tag{25}$$

where, $\boldsymbol{q}_{\Gamma}$ denotes the flow across the element boundary, and the unit outward normal vector is represented as $\boldsymbol{n}_{\mathrm{e}}$. The vector $\boldsymbol{N_P}$ corresponds to the shape functions, $\boldsymbol{B_P} = \nabla \boldsymbol{N_P}$, while $\boldsymbol{P}_{\mathrm{e}}$ signifies the pressure DOF associated with the element e. Under the condition that both outlet pressure $p_{\mathrm{out}}$ and boundary flux $\boldsymbol{q}_{\Gamma}$ are set to zero, the equilibrium equation at the element level simplifies to:

$$\boldsymbol{K}_{\boldsymbol{P}}^{\mathrm{e}} \boldsymbol{P}_{\mathrm{e}} + \boldsymbol{K}_{\boldsymbol{DP}}^{\mathrm{e}} \boldsymbol{P} = \boldsymbol{A}_{\mathrm{e}} \boldsymbol{P}_{\mathrm{e}} = \boldsymbol{0}, \tag{26}$$

where, $\boldsymbol{K}_{\boldsymbol{P}}^{\mathrm{e}} = \int_{\Omega_{\mathrm{e}}} \left( K \boldsymbol{B}_{\boldsymbol{P}}^{\mathrm{T}} \boldsymbol{B}_{\boldsymbol{P}} \right) \mathrm{d}V$ and $\boldsymbol{K}_{\boldsymbol{DP}}^{\mathrm{e}} = \int_{\Omega_{\mathrm{e}}} \left( D \boldsymbol{N}_{\boldsymbol{P}}^{\mathrm{T}} \boldsymbol{N}_{\boldsymbol{P}} \right) \mathrm{d}V$ denote the element flow stiffness matrix and element drainage matrix. $\boldsymbol{A}_{\mathrm{e}}$ is the element flow vector. By assembling the contributions from all elements, the global equilibrium equation for the structure can be obtained as:

$$\boldsymbol{A}\boldsymbol{P} = \boldsymbol{0}, \tag{27}$$

where, $\boldsymbol{A}$ and $\boldsymbol{P}$ denote the global flow matrix and the global pressure vector. By solving Eq. (27), the continuous pressure distribution associated with the design variables can be obtained without the need for explicit boundary tracking.

#### *4.4.2 Conversion of pressure field to nodal loads*

To ensure compatibility between continuous pressure distribution and discretized finite element formulation, it is essential to convert the pressure load into a nodal force vector that aligns with the structural DOF. Following the methodology proposed by Kumar et al. (2020) the boundary pressure field $p$ can be transformed into an equivalent body force $\boldsymbol{b}$ through the equilibrium relation:

$$\boldsymbol{b}\mathrm{d}V = -\nabla p \mathrm{d}V. \tag{28}$$

Therefore, the virtual work done by the body force within an element can be expressed as:

$$\delta W_{\mathrm{e}} = \int_{\Omega_{\mathrm{e}}} \delta \boldsymbol{u}^{\mathrm{T}} \boldsymbol{b} \mathrm{d}V = \delta \boldsymbol{d}_{\mathrm{e}}^{\mathrm{T}} \int_{\Omega_{\mathrm{e}}} \boldsymbol{N}_{\boldsymbol{u}}^{\mathrm{T}} \boldsymbol{b} \mathrm{d}V, \tag{29}$$

where, $\boldsymbol{N}_{\boldsymbol{u}}$ denotes the shape function vector associated with the displacement field, and $\boldsymbol{d}_{\mathrm{e}}$ represents elemental DOF vector. Based on the principle of virtual work consistency and Eq. (29), the

`

relationship between the elemental DOF vector and the elemental pressure DOF can be derived as follows:

$$\boldsymbol{F}_{\mathrm{e}} = \int_{\Omega_{\mathrm{e}}} \boldsymbol{N}_{\boldsymbol{u}}^{\mathrm{T}} \left(-\nabla p\right) \mathrm{d}V = -\underbrace{\left(\int_{\Omega_{\mathrm{e}}} \boldsymbol{N}_{\boldsymbol{u}}^{\mathrm{T}} \boldsymbol{B}_{\boldsymbol{P}} \mathrm{d}V\right)}_{\boldsymbol{T}_{\mathrm{e}}} \boldsymbol{P}_{\mathrm{e}}, \tag{30}$$

where, $\boldsymbol{T}_{\mathrm{e}}$ denotes the transformation matrix for element e, and $\boldsymbol{F}_{\mathrm{e}}$ represents the corresponding elemental nodal force vector. By applying the transformation $\boldsymbol{T}_{\mathrm{e}}$ to each element and assembling all elemental force vectors, the global nodal force vector $\boldsymbol{F}_{\mathrm{global}}$ is obtained. This leads to the final discretized structural equilibrium equation, given by:

$$\boldsymbol{K}\left(\boldsymbol{\rho}\right)\boldsymbol{u} = \boldsymbol{F}_{\mathrm{global}} = -\boldsymbol{T}\boldsymbol{P}, \tag{31}$$

where, $\boldsymbol{K}(\boldsymbol{\rho})$ denotes the global stiffness matrix of the structure, and $\boldsymbol{T}$ represents the global transformation matrix assembled from $\boldsymbol{T}_{\mathrm{e}}$.

#### *4.4.3 Topology optimization with design-dependent fluid pressure loads*

This section explores the application of DPIKAN-TO in topology optimization problems that involve design-dependent fluid pressure loads. To illustrate the feasibility of DPIKAN-TO, two distinct optimization problems under fluid pressure loads are investigated: structural design and compliant mechanism design. In most conventional topology optimization problems, external loads are typically assumed to remain constant. However, in numerous practical situations, the location, magnitude, and direction of these applied loads evolve concurrently with material distribution, thereby introducing additional challenges to the optimization process. To address these challenges, we adopts the method proposed by Kumar et al.(2020) to resolve the continuous pressure field and transform it into nodal loads. By integrating this fluid–structure coupling mechanism into DPIKAN-TO, we extend its applicability to topology optimization problems involving design-dependent fluid pressure loads. The mathematical formulation for the compliance minimization of a rigid structure subjected to design-dependent pressure loading is given as follows:

$$
\begin{aligned}
\min \quad & C = \boldsymbol{u}^{\mathrm{T}}\boldsymbol{K}\boldsymbol{u} = 2\mathrm{SE} \\
\text{s.t.} \quad & \boldsymbol{AP} = \boldsymbol{0} \\
& \boldsymbol{Ku} = -\boldsymbol{TP} \\
& f_{\mathrm{v}}(\boldsymbol{\rho}) = \frac{\sum_{i\in\mathbb{N}_{\mathrm{e}}} \rho_i v_i}{V} = V_f \\
& 0 \le \rho_i \le 1, \quad i \in \mathbb{N}_{\mathrm{e}},
\end{aligned} \tag{32}
$$

where SE represents the structural strain energy. It is crucial to note that, since the conversion of pressure loads into equivalent nodal forces occurs outside the framework of the neural network, the sensitivity of the fluid pressure field concerning design variables must be explicitly computed rather than implicitly learned by the network (Chandrasekhar and Suresh 2021). Compared to other methods for pressure-loaded structures, the approach proposed by Kumar et al. (2023) explicitly incorporates load sensitivity, which can affect both the structural shape and topological configuration during the optimization process. Therefore, the loss function of the s-HRKAN for minimum compliance design of structures subjected to pressure loading can be formulated as:

$$
\arg\min L_{\mathrm{obj}}(\boldsymbol{\theta}) = \frac{\boldsymbol{u}^{\mathrm{T}}\boldsymbol{K}\boldsymbol{u}}{C_0} + \lambda_{lst}\boldsymbol{AP} + \alpha\left(\frac{f_{\mathrm{v}}(\boldsymbol{\rho})}{V_f} - 1\right)^2, \tag{33}
$$

where $\lambda_{\mathrm{lst}}$ is computed explicitly as $\lambda_{\mathrm{lst}}^{\mathrm{T}} = 2\boldsymbol{u}^{\mathrm{T}}\boldsymbol{T}\boldsymbol{A}^{-1}$.To verify the design feasibility of DPIKAN-TO, its results are compared with those obtained using the SIMP method implemented in the TOPress code by Kumar et al. (2023).

#### *4.4.4 Compliance minimization under Design-Dependent Fluid Pressure Loads*

This section evaluates the proposed DPIKAN-TO framework through two benchmark cases: the arch structure examined by Hammer and Olhoff (2000), and the piston structure investigated by Bourdin and Chambolle (2003). The initial design domains for the two cases are illustrated in Fig. 15. The arch structure is modeled as a deep beam with dimensions of 0.8×0.4, discretized into $80 \times 40$ elements ($L_e$ = 0.01). Meanwhile, the piston structure utilizes a domain measuring, comprising $120 \times 40$ elements ($L_e$ = 0.01). A unit thickness ($t$ = 1) is assumed for both structures. The boundary pressure is set to $p$=1, with zero pressure($p_0$) on the remaining edges. The two cases are subject to volume fraction constraints of $V_f$ = 0.25 and $V_f$ = 0.5, respectively. The pressure field responses of the structure are illustrated in Fig. 16. In this context, “lst” denotes whether load sensitivity is incorporated into the objective function sensitivity analysis. Specifically, lst = 1 means the load sensitivity is considered, while lst = 0 means it is omitted.

The Darcy flow–drainage model is configured with projection parameters $\eta_k = \eta_h = 0.25$ and $\beta_k = \beta_h = 10$ for both the flow and drainage coefficients. The flow coefficients are established as $k_v = 1$ for voids and $k_s = 1\times10^{-7}$ for solids. The pressure penetration depth and ratio are defined as $\Delta s = 2$ and r = 0.1. Material properties, topology optimization settings, and neural network hyperparameters remain consistent with those outlined in Section 4.1.

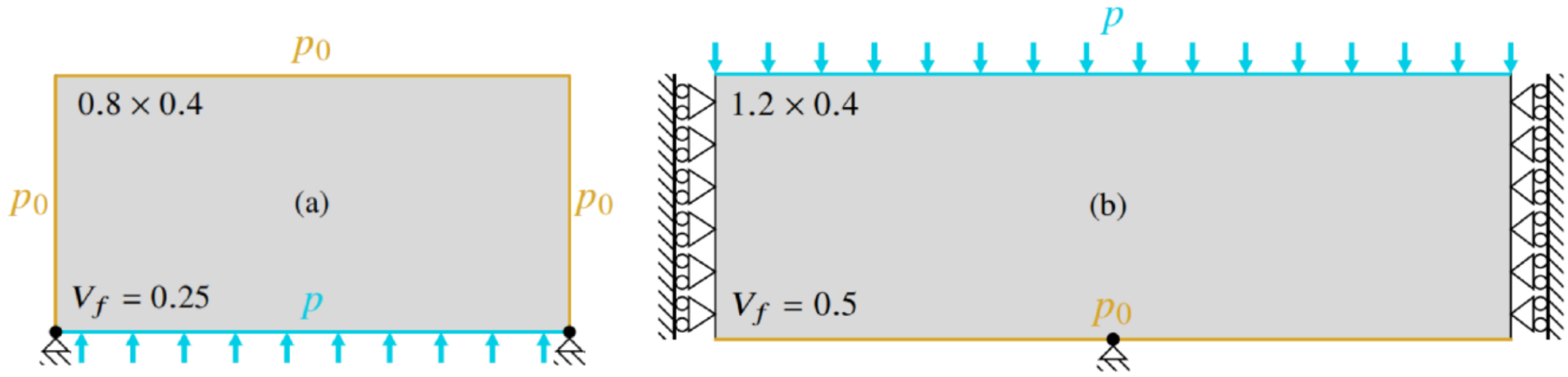

Fig. 15 Initial design domains for compliance minimization under design-dependent fluid pressure loads: (a) Arch structure; (b) Piston structure.

Based on the final topologies, pressure field responses, and convergence results presented in Fig. 16 and Fig. 17, several observations can be drawn. In the case of the arch structure, the design generated by DPIKAN-TO closely aligns with that achieved through the SIMP method. Both designs demonstrate higher stiffness compared to those that without load sensitivity, which is attributed to the incorporation of load sensitivity which effectively reduces the height of the resulting arch structure. Furthermore, differences are evident between the DPIKAN-TO and SIMP designs in the piston case. The proposed method yields a topology characterized by fewer internal voids. When accounting for load sensitivity, the final compliance differences between the designs generated by DPIKAN-TO and those from the SIMP method are only approximately 0.18% and 0.47%. This demonstrates that DPIKAN-TO is capable of producing designs comparable to those obtained by SIMP.

lst=1 lst=0

0.0

Pressure

1.0

(a) Iter: 88, Obj:33.21 via DPIKAN-TO

(b) Iter: 105, Obj:35.23 via DPIKAN-TO

(c) Iter: 200, Obj:33.15 via SIMP

(d) Iter: 200, Obj:34.78 via SIMP

**Fig. 16** Final topologies and pressure responses of the arch structure with load sensitivity (lst = 1) and without load sensitivity (lst = 0).

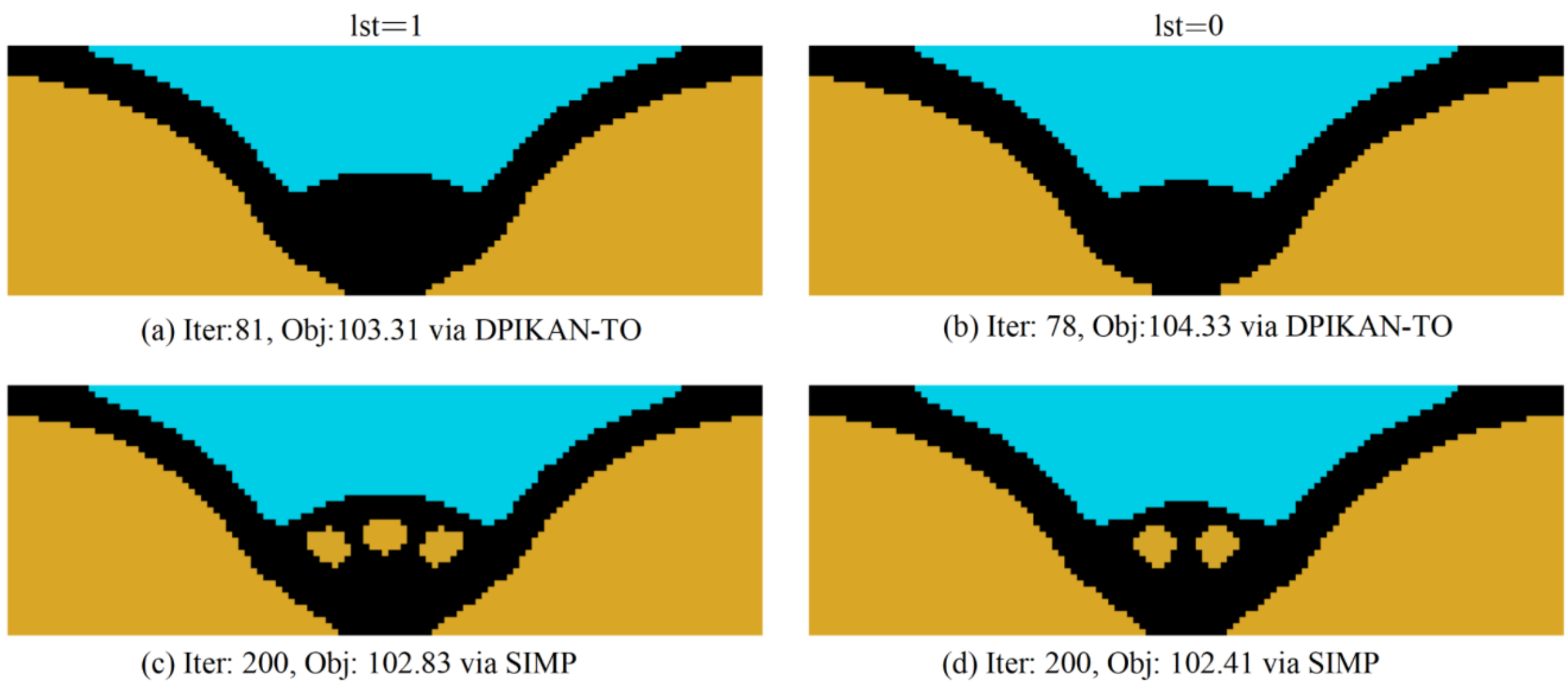

**Fig. 17** Final topologies and pressure responses of the piston structure with load sensitivity and without load sensitivity.

### *4.4.5 Compliant mechanism under design-dependent fluid pressure loads*

This section examines the application of DPIKAN-TO in the topology optimization of compliant mechanisms. A compliant inverter mechanism is employed as an illustrative example, in which the structure is designed to deform in a direction opposite to the applied pressure load. Typically, the design objective is to maximize output displacement while ensuring adequate structural stiffness (Saxena and Ananthasuresh 2000). Therefore, based on the multi-criteria compliance-stiffness formula (Saxena and Ananthasuresh 2000;Frecker et al. 1997), we employs mutual strain energy (MSE) to characterize both the output deformation of the structure and its internal energy. Consequently, we design an optimal topology for the inverter mechanism with the objectives of maximizing structural stiffness and output deformation. The mathematical formulation for the design problem concerning compliant inverter mechanisms subjected to pressure loads is presented as follows (Kumar et al. 2020):

$$\begin{aligned} \min \ & f_0^{\mathrm{CM}}(\boldsymbol{u},\boldsymbol{v},\boldsymbol{\rho}) = -\frac{\mathrm{MSE}}{2\mathrm{SE}} \\ \text{s.t.} \ & \boldsymbol{AP} = \boldsymbol{0} \\ & \boldsymbol{Ku} = -\boldsymbol{TP} \\ & \boldsymbol{Kv} = \boldsymbol{F}_{\mathrm{d}} \\ & f_{\mathrm{v}}(\boldsymbol{\rho}) = \frac{\sum_{i\in\mathbb{N}_{\mathrm{e}}} \rho_i v_i}{V} = V_f \\ & 0 \le \rho_i \le 1, \quad i \in \mathbb{N}_{\mathrm{e}}, \end{aligned} \tag{34}$$

where, $f_0^{\mathrm{CM}}$ is the multi-criteria objective function, and $\boldsymbol{F}_{\mathrm{d}}$ is the unit virtual force vector in the same direction as the output position displacement. The vector $\boldsymbol{v}$ represents the structural virtual displacement caused by the unit virtual force, while MSE = $\boldsymbol{v}^{\mathrm{T}}\boldsymbol{Ku}$ represents the mutual strain energy.

By leveraging the symmetry in the design, we focus exclusively on optimizing the upper half of the compliant inverter, as illustrated in Fig. 18. The dimensions of the design domain are set at 1.0 × 0.5 and discretized into a grid comprising 100 × 50 elements ($L_{\mathrm{e}}$ = 0.01). The thickness is set at $t$ = 1, and the volume fraction constraint value $V_f$ is set at 0.3. The projection parameters in the element flow and drainage coefficient are set as $\eta_{\mathrm{k}} = \eta_{\mathrm{h}} = 0.6$, $\beta_{\mathrm{k}} = \beta_{\mathrm{h}} = 10$. The settings for material properties, topological parameters, and the remaining Darcy flow-drainage model parameters are consistent with those in the section 4.4.4 dealing with the rigid structure design problem under pressure loading.

In addition, a spring with a stiffness of $k_{out} = 1.5\times10^{-3}$ was incorporated at the output position of the structure to model the reaction force generated at this location. To ensure the stability of the training process, we adjust the initial learning rate of s-HRKAN within the DPIKAN-TO network structure to $1\times10^{-3}$. Simultaneously, the d-HRKAN adopts an architecture of $NN_{d\text{-}HRKAN}$ = [2, 40, 40, 2] with unchanged hyperparameters, and its early stopping iteration limit is doubled. The structural virtual displacement induced by the unit virtual force vector applied at the output port is directly obtained through FEA.

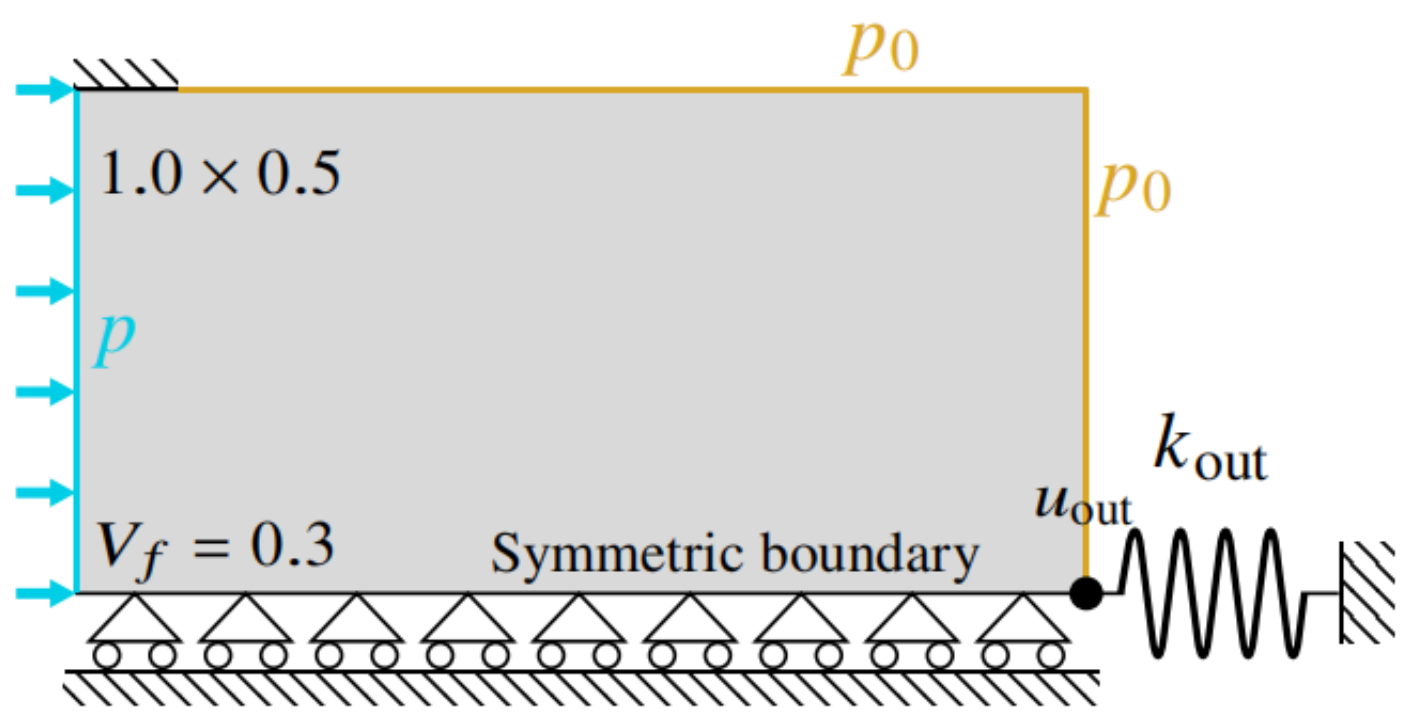

**Fig. 18** Initial design domains for compliant inverter mechanism under design-dependent fluid pressure loads.

By examining the design results of the compliant inverter mechanism obtained through DPIKAN-TO and SIMP, as illustrated in Fig. 19, it is evident that there are notable differences in the topologies produced by these two methods. However, both approaches exhibit hinge characteristics. This phenomenon can be attributed to the relatively low spring stiffness ($k_{out}$) employed in the analysis. The presence of hinges can improve the flexibility of compliant mechanisms, facilitating a more efficient transmission of energy to the output position via these pathways. However, this enhancement also leads to a significant reduction in the overall structural stiffness (Bruns and Tortorelli 2001). Due to the differences in the generated designs, the design produced by DPIKNA-TO exhibits fewer hinge features, resulting in higher structural stiffness compared to the SIMP design. However, this characteristic also leads to a reduced displacement at the output end of the DPIKNA-TO design. This observation demonstrates the feasibility of DPIKNA-TO in addressing complex topology optimization problems involving pressure loads.

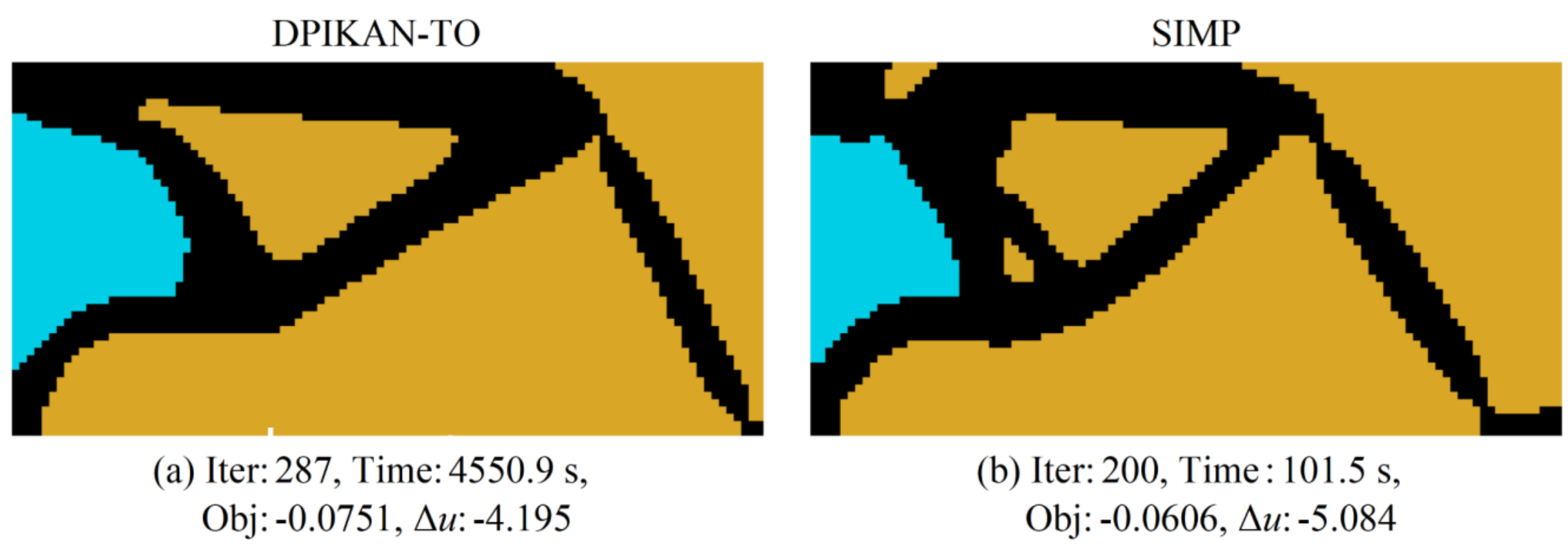

**Fig. 19** Final topologies and pressure responses of the compliant inverter mechanism with load sensitivity and without load sensitivity.

### *4.5 Three-Dimensional topology optimization problems*

This section extends the DPIKAN-TO framework to 3D topology optimization and demonstrates its feasibility and effectiveness through a representative compliance minimization example. As illustrated in Fig. 20, the optimization target is a cantilever beam with dimensions of 4×2×2, and its design domain is discretized into 40×20×20 elements ($L_e$ = 0.1). The magnitude of the distributed load at the boundary is established as $f$ = 1, which is uniformly applied across 20 sampling points. The volume fraction constraint is set to $V_f$ = 0.5. The material properties and topology optimization parameters are consistent with those presented in Section 4.1.

To account for the non-negligible stress and strain in the z-direction in 3D structures, an additional neuron is incorporated into the input layer to encode the z-coordinate. Furthermore, the output layer is expanded to facilitate predictions of displacement along the z-direction. To accommodate the extension of input dimensions from 2D to 3D and to enhance the network's capacity for capturing high-order couplings among variables, we augment the original s-HRKAN by incorporating an additional HRKAN layer. This modification results in a four-layer network architecture (e.g., $NN_{s\text{-}HRKAN}$ = [3, 40, 40, 40, 1]), while maintaining the original hyperparameter configuration. The designs generated through DPIKAN-TO are subsequently compared with those obtained using the SIMP method coded by Liu and Tovar (2014). For visualization purposes, the final topologies produced by both SIMP and DPIKAN-TO are processed using a unified isosurface threshold ($\rho$ = 0.5) in Paraview. It is noted that, to reduce the increased compliance caused by asymmetry from analysis errors, symmetry constraints are enforced within the DPIKAN-TO framework.

By comparing the topological designs and convergence results of the 3D cantilever beam generated by SIMP and DPIKAN-TO with different s-HRKAN architectures, which is presented in Fig. 20. It is evident that when applying the three-layer network architecture s-HRKAN directly to the 3D topology optimization problem (as shown in Fig. 20 (c)), there is a significant increase in input space dimensionality. This expansion complicates the mapping relationship necessary for fitting, thereby limiting the network's expressive capability and hindering its ability to capture intricate features within 3D topology design. The combined analysis of Fig. 20 and Fig. 21 indicates that the four-layer s-HRKAN employed in DPIKAN-TO demonstrates superior design quality compared to its three-layer counterpart. This enhanced model captures finer structural details and achieves a more rapid convergence of compliance. As illustrated in Fig. 20 (b) and 20(d), the final topology produced by the four-layer network closely resembles that generated by SIMP, with both exhibiting intricate internal voids. Moreover, the compliance of this configuration is approximately 4.74% lower than that of the SIMP result, signifying an improvement in structural stiffness. The performance advantage is attributed not only to the topological differences between the designs but also to the lower proportion of gray (intermediate-density) elements in the designs obtained from DPIKAN-TO. This reduction enhances both design clarity and manufacturability.

DPIKAN-TO necessitates significantly fewer iterations compared to the SIMP method, which can be ascribed to both the convergence behavior of HRKAN. It is important to note that while increasing the network depth to four layers facilitates faster convergence of compliance in DPIKAN-TO, excessively wide layers (e.g., Fig. 20 (f)) result in longer computational times due to the elevated cost per iteration. Furthermore, overly wide architectures tend to obscure fine structural details in the final topology. Although this challenge can be addressed by reducing the learning rate of the s-HRKAN, such an adjustment will further exacerbate the computational burden.

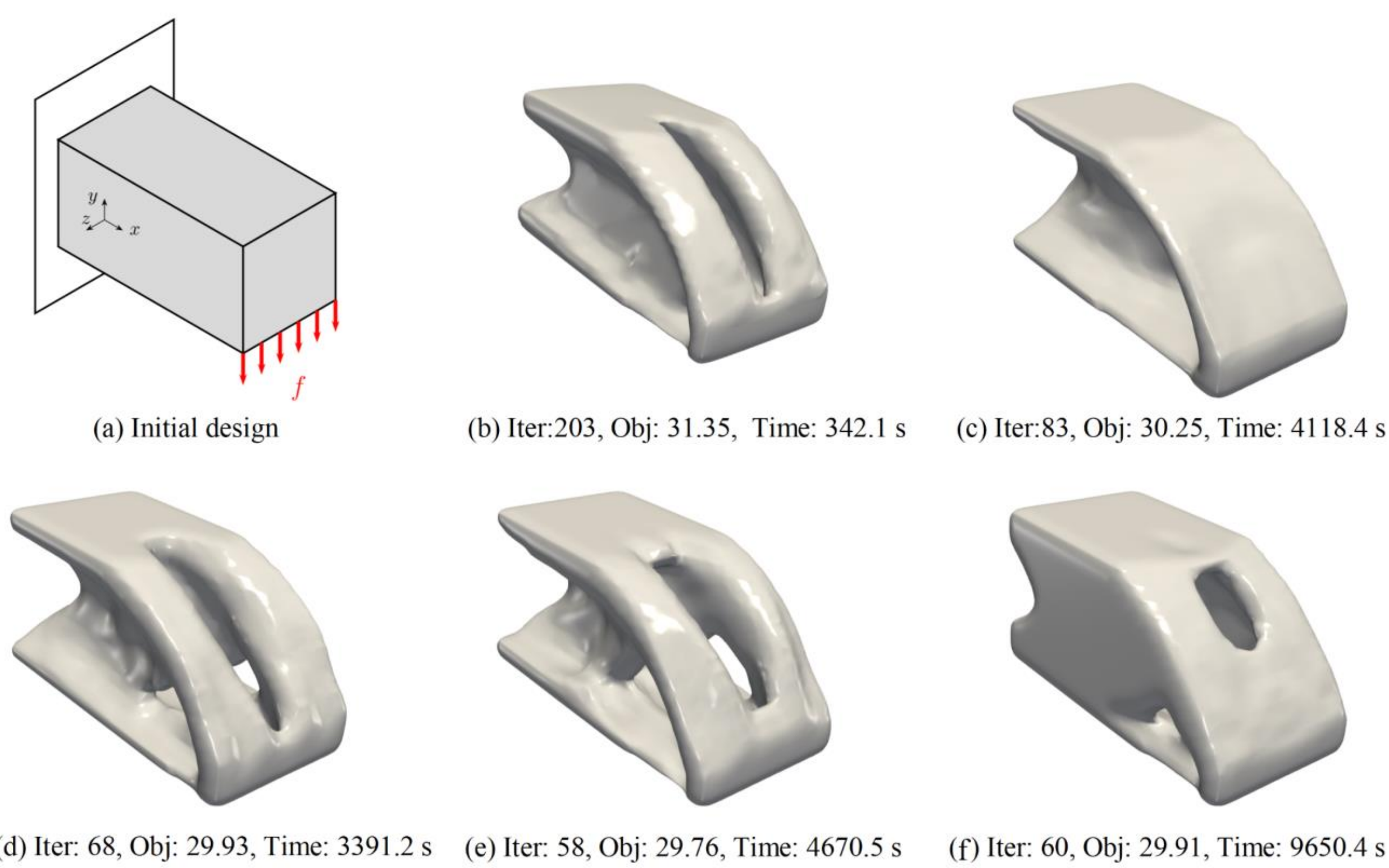

**Fig. 20** Topology and convergence results for 3D cantilever compliance minimization using SIMP and DPIKAN-TO with different s-HRKAN architectures: (a) Initial design domain. (b) SIMP result. (c) DPIKAN-TO with [3, 40, 40, 1]. (d) DPIKAN-TO with [3, 40, 40, 40, 1]. (e) DPIKAN-TO with [3, 80, 80, 80, 1]. (f) DPIKAN-TO with [3, 120, 120, 120, 1].

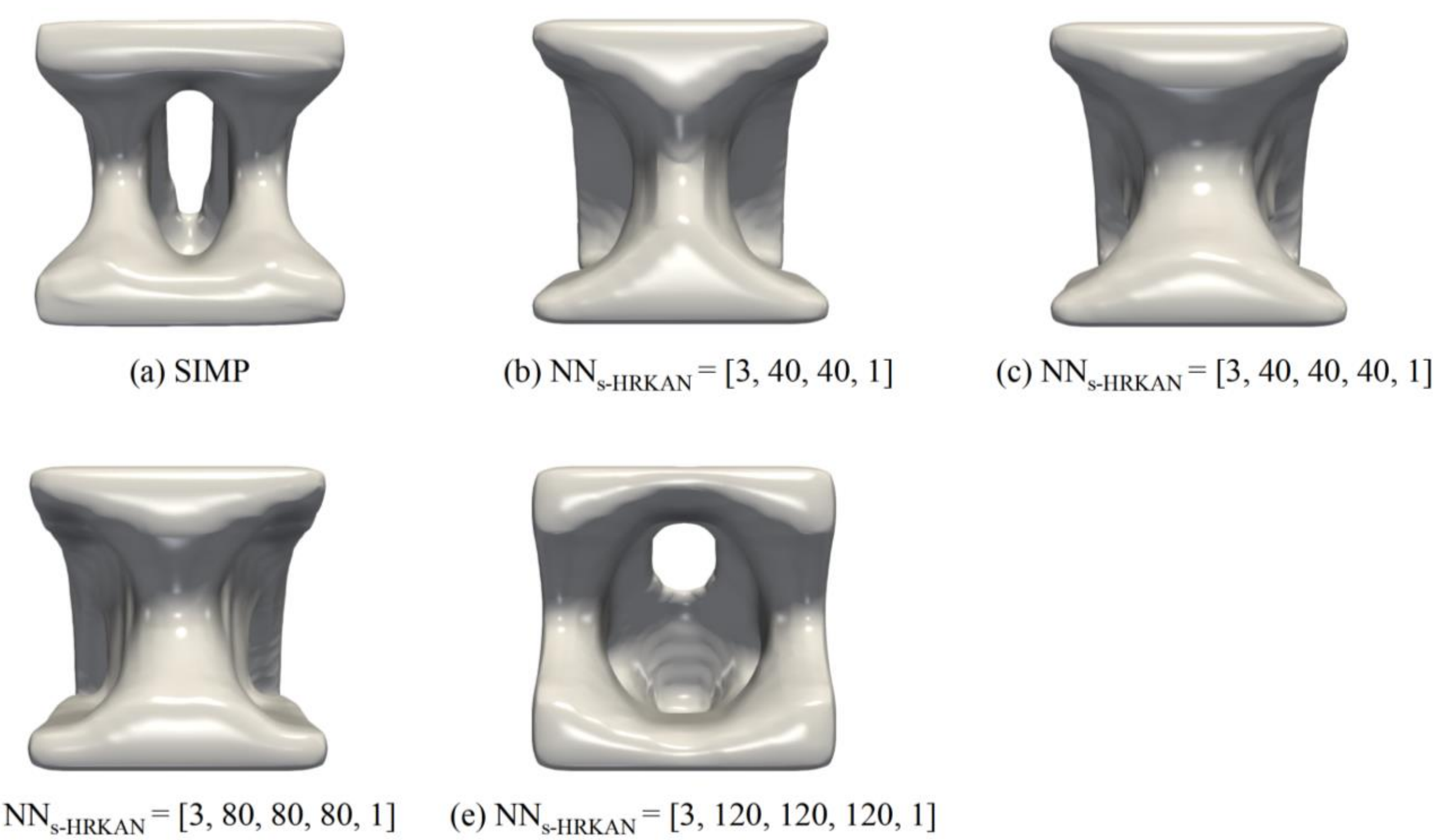

**Fig. 21** Rear views of the 3D cantilever topologies in Fig. 20: SIMP and DPIKAN-TO with different s-HRKAN architectures.

### *4.6 Discussion on computational cost and applicability*

Although the proposed DPIKAN-TO framework demonstrates improved convergence behavior and enhanced topology representation capability compared with conventional PINN-based topology optimization approaches, its computational cost remains substantially higher than that of classical FEM-based methods for standard linear elastic benchmark problems. This discrepancy primarily originates from the repeated neural training process required for PDE approximation. In conventional finite element analysis, the equilibrium equation can be solved directly through stiffness matrix assembly and linear system solution. By contrast, d-HRKAN requires iterative neural optimization to approximate the equilibrium response field during each topology update step, resulting in significantly higher computational overhead.

Therefore, the primary significance of the proposed framework does not lie in replacing FEM-based topology optimization for classical low-complexity linear elastic problems. Instead, its potential is expected to become more evident in problems involving strongly coupled multiphysics PDEs, irregular geometries, nonlinear constitutive behavior, inverse design, and differentiable end-to-end optimization requirements. In such scenarios, conventional FEM-based optimization pipelines often involve heterogeneous numerical modules and explicit sensitivity derivations, which are difficult to integrate into a unified differentiable framework. By contrast, the proposed dual-surrogate paradigm provides a unified neural representation for both equilibrium evolution and topology evolution, thereby offering improved extensibility toward complex PDE-constrained optimization systems.

## 5. Conclusions

This work proposes a Dual Physics-Informed Kolmogorov–Arnold Network–based Continuum Topology Optimization (DPIKAN-TO) framework for solving a broad class of continuum topology optimization problems, including linear elastic structures, pressure-dependent loading conditions, stress-constrained designs, and compliant mechanisms. Within the proposed framework, the DEM-PINN and S-PINN are coupled with a Hierarchical Residual Kolmogorov–Arnold Network (HRKAN) to form the d-HRKAN and s-HRKAN modules, respectively. These modules are responsible for structural analysis (i.e., solving the governing partial differential equations) and for updating the design variables during the topology optimization process. Numerical investigations demonstrate that, for two-dimensional compliance minimization problems, DPIKAN-TO achieves significantly improved computational efficiency compared to CPINNTO (Jeong et al. 2023), while partially mitigating spectral bias and producing topological designs comparable to those obtained using the SIMP method. For more complex problems involving stress constraints and compliant mechanism design, DPIKAN-TO consistently generates solutions that satisfy optimality conditions similar to those achieved by SIMP. Furthermore, in challenging three-dimensional topology optimization problems—where existing PINN-based approaches often struggle—DPIKAN-TO is capable of producing results comparable to SIMP while requiring substantially fewer optimization iterations.

**Application of the proposed approach:** The broad applicability of DPIKAN-TO has been preliminarily demonstrated through a diverse set of continuum topology optimization benchmark problems. Meanwhile, by exploiting the localized and high-dimensional feature representation capability of HRKAN, the proposed framework can better capture sharp topology interfaces, localized structural features, and rapidly varying sensitivity responses. These characteristics make DPIKAN-TO particularly promising for high-reliability structural design problems involving stress concentration, complex loading conditions, and multiphysics coupling. Therefore, the proposed dual physics-informed surrogate framework provides a potential foundation for extending neural-network-based topology optimization toward more complex engineering design scenarios.

**Limitation of the Study:** Several limitations remain in the current DPIKAN-TO framework. First, although DPIKAN-TO requires fewer optimization iterations than the SIMP method in linear elastic structural optimization problems, its total computational cost is still higher than that of classical FEM-based approaches. Second, the optimization performance remains sensitive to the architectures and hyperparameters of both d-HRKAN and s-HRKAN, which may affect convergence stability and final topology quality. Third, mesh dependency is not completely eliminated, since Gaussian quadrature and discretized sampling are still employed in the loss evaluation. Nevertheless, the computational advantage of DPIKAN-TO may become more evident in nonlinear or strongly coupled problems, where conventional FEA often requires multi-step iterative solution procedures and may encounter convergence difficulties. Fourth, the extension to large-scale three-dimensional and strongly coupled multiphysics problems still requires further improvements in computational efficiency and memory management.

**Future recommendations:** Future research will focus on improving the scalability, robustness, and computational efficiency of the DPIKAN-TO framework. Promising directions include integrating neural operators to accelerate PDE-governed equilibrium-field approximation, introducing adaptive sampling strategies to improve accuracy in regions with sharp density transitions and stress concentrations, and developing multi-resolution surrogate representations to capture both global structural layouts and localized high-frequency features. In addition, transfer learning and pretraining strategies may be explored to enhance convergence stability when extending the framework to large-scale three-dimensional and strongly coupled multiphysics topology optimization problems. These improvements are expected to further strengthen the applicability of DPIKAN-TO in complex engineering design scenarios.

## CRediT authorship contribution statement

**Junyuan Zhang:** Writing – review & editing, Writing – original draft, Visualization, Validation, Software, Methodology, Investigation, Data curation. **Jing Cao:** Writing – review & editing, Methodology. **Abdullah Dawar:** Writing – review & editing, Supervision. **Kun Cai:** Conceptualization, Supervision, Writing – review & editing, Formal analysis. **QingHua Qin:** Supervision, Writing – review & editing.

## Data availability

No data was used for the research described in the article.

## Declaration of competing interest

The authors declare that they have no known competing financial interests or personal relationships that could have appeared to influence the work reported in this paper.

## Appendix A. Efficiency Comparison: DPIKAN-TO vs. CPINNTO

This section compares the computational efficiency of DPIKAN-TO and CPINNTO using four standard 2D topology optimization benchmarks from Jeong et al. (2023). Both methods use the same material properties and convergence criteria, Specifically, while satisfying the volume constraint and ensuring that the minimum convergence error does not exceed the threshold $\tau = 0.08$. As illustrated in Tables A.1 and A.2, DPIKAN-TO exhibits a significant enhancement in convergence efficiency when compared to CPINNTO across the four 2D benchmark problems. It is worth noting that the CPINNTO network contains 26,162 trainable parameters, whereas the DPIKAN-TO involves 30,837 trainable parameters in total.

**Table A.1**

Computational efficiency of CPINNTO for 2-D benchmark problems from Jeong et al. (2023).

| Case | Average CPU time (seconds) | | | Total CPU time(seconds) | | | |
|---|---|---|---|---|---|---|---|
| | SIMP | SIMP + Heaviside | CPINNTO | SIMP | SIMP + Heaviside | CPINNTO | $t_{CPINNTO}/t_{SIMP+Heaviside}$ |
| (a) | 0.35 | 0.35 | 74 | 53.95 | 492.63 | 16872 | 34.249 |
| (b) | 0.31 | 0.33 | 74 | 22.67 | 338.35 | 34558 | 102.137 |
| (c) | 0.31 | 0.32 | 83 | 13.18 | 257.04 | 9960 | 38.749 |
| (d) | 0.32 | 0.32 | 77 | 8.01 | 225.14 | 17171 | 76.268 |

**Table A.2**

Computational efficiency of DPIKAN-TO on 2D benchmark problems.

| Case | Average CPU time (seconds) | | | Total CPU time(seconds) | | | |
|---|---|---|---|---|---|---|---|
| | SIMP | SIMP+Heaviside | DPIKAN-TO | SIMP | SIMP+Heaviside | DPIKAN-TO | $t_{DPIKANTO}/t_{SIMP+Heaviside}$ |
| (a) | 0.035 | 0.041 | 4.570 | 5.469 | 26.115 | 484.452 | 18.551 |
| (b) | 0.037 | 0.037 | 4.499 | 2.529 | 18.192 | 499.422 | 27.453 |
| (c) | 0.048 | 0.041 | 4.465 | 1.159 | 19.734 | 281.265 | 15.253 |
| (d) | 0.048 | 0.039 | 4.398 | 1.197 | 18.531 | 1147.944 | 61.948 |

## References

Aage, N., Andreassen, E., & Lazarov, B. S. (2015). Topology optimization using PETSc: An easy-to-use, fully parallel, open source topology optimization framework. Structural and Multidisciplinary Optimization, *51*(3), 565-572, doi:https://doi.org/10.1007/s00158-014-1157-0.

Amir, O., Sigmund, O., Lazarov, B. S., & Schevenels, M. (2012). Efficient reanalysis techniques for robust topology

optimization. Computer Methods in Applied Mechanics Engineering, *245*, 217-231, doi:https://doi.org/10.1016/j.cma.2012.07.008.

Andreassen, E., Clausen, A., Schevenels, M., Lazarov, B. S., & Sigmund, O. (2011). Efficient topology optimization in MATLAB using 88 lines of code. Structural and Multidisciplinary Optimization, *43*, 1-16, doi:https://doi.org/10.1007/s00158-010-0594-7.

Bendsøe, M. P. (1989). Optimal shape design as a material distribution problem. Structural and Multidisciplinary Optimization, *1*(4), 193-202, doi:https://doi.org/10.1007/BF01650949.

Bendsøe, M. P. (2008). Topology optimization. In *Encyclopedia of Optimization* (pp. 3928-3929). Boston,MA: Springer.

Bendsøe, M. P., & Sigmund, O. (1999). Material interpolation schemes in topology optimization. Archive of Applied Mechanics, *69*(9), 635-654, doi:https://doi.org/10.1007/s004190050248.

Bourdin, B., & Chambolle, A. (2003). Design-dependent loads in topology optimization. ESAIM: Control, Optimisation and Calculus of Variations, *9*, 19-48, doi:https://doi.org/10.1051/cocv:2002070.

Bruns, T. E., & Tortorelli, D. A. (2001). Topology optimization of non-linear elastic structures and compliant mechanisms. Computer Methods in Applied Mechanics Engineering, *190*(26-27), 3443-3459, doi:https://doi.org/10.1016/S0045-7825(00)00278-4.

Cavazzuti, M., Baldini, A., Bertocchi, E., Costi, D., Torricelli, E., et al. (2011). High performance automotive chassis design: a topology optimization based approach. Structural and Multidisciplinary Optimization, *44*(1), 45-56, doi:https://doi.org/10.1007/s00158-010-0578-7.

Chadha, K. S., & Kumar, P. (2025). PyTOaCNN: Topology optimization using an adaptive convolutional neural network in Python. Soft Computing, 1-21.

Chandrasekhar, A., & Suresh, K. (2021). TOuNN: Topology optimization using neural networks. Structural and Multidisciplinary Optimization, *63*(3), 1135-1149, doi:https://doi.org/10.1007/s00158-020-02748-4.

Chen, A., Cai, K., Zhao, Z.-L., Zhou, Y., Xia, L., et al. (2021). Controlling the maximum first principal stress in topology optimization. Structural and Multidisciplinary Optimization, *63*(1), 327-339, doi:https://doi.org/10.1007/s00158-020-02701-5.

Deng, H., Vulimiri, P. S., & To, A. C. (2022). An efficient 146-line 3D sensitivity analysis code of stress-based topology optimization written in MATLAB. Optimization and Engineering, *23*(3), 1733-1757, doi:https://doi.org/10.1007/s11081-021-09675-3.

Farea, A., & Celebi, M. S. (2025). Learnable activation functions in physics-informed neural networks for solving partial differential equations. Computer Physics Communications, 109753, doi:https://doi.org/10.1016/j.cpc.2025.109753.

Frecker, M., Ananthasuresh, G., Nishiwaki, S., Kikuchi, N., & Kota, S. (1997). Topological synthesis of compliant mechanisms using multi-criteria optimization. Journal of Mechanical Design, doi:https://doi.org/10.1115/1.2826242.

Gogu, C. (2015). Improving the efficiency of large scale topology optimization through on-the-fly reduced order model construction. International Journal for Numerical Methods in Engineering, *101*(4), 281-304, doi:https://doi.org/10.1002/nme.4797.

Guest, J. K., Prévost, J. H., & Belytschko, T. (2004). Achieving minimum length scale in topology optimization using nodal design variables and projection functions. International Journal for Numerical Methods in Engineering, *61*(2), 238-254,

doi:https://doi.org/10.1002/nme.1064.

Guo, X., Zhang, W., Zhang, J., & Yuan, J. (2016). Explicit structural topology optimization based on moving morphable components (MMC) with curved skeletons. Computer Methods in Applied Mechanics Engineering, *310*, 711-748, doi:https://doi.org/10.1016/j.cma.2016.07.018.

Hammer, V. B., & Olhoff, N. (2000). Topology optimization of continuum structures subjected to pressure loading. Structural and Multidisciplinary Optimization, *19*(2), 85-92, doi:https://doi.org/10.1007/s001580050088.

He, J., Abueidda, D., Al-Rub, R. A., Koric, S., & Jasiuk, I. (2023a). A deep learning energy-based method for classical elastoplasticity. International Journal of Plasticity, *162*, 103531, doi:https://doi.org/10.1016/j.ijplas.2023.103531.

He, J., Chadha, C., Kushwaha, S., Koric, S., Abueidda, D., et al. (2023b). Deep energy method in topology optimization applications. Acta Mechanica, *234*(4), 1365-1379, doi:https://doi.org/10.1007/s00707-022-03449-3.

Holmberg, E., Torstenfelt, B., & Klarbring, A. (2013). Stress constrained topology optimization. Structural and Multidisciplinary Optimization, *48*(1), 33-47, doi:https://doi.org/10.1007/s00158-012-0880-7.

Huang, X., & Xie, Y.-M. (2010). A further review of ESO type methods for topology optimization. Structural and Multidisciplinary Optimization, *41*(5), 671-683, doi:https://doi.org/10.1007/s00158-010-0487-9.

Ioffe, S., & Szegedy, C. Batch normalization: Accelerating deep network training by reducing internal covariate shift. In *International conference on machine learning, 2015* (pp. 448-456): pmlr. doi:https://doi.org/10.48550/arXiv.1502.03167.

Jeong, H., Batuwatta-Gamage, C., Bai, J., Rathnayaka, C., Zhou, Y., et al. (2025). An advanced physics-informed neural network-based framework for nonlinear and complex topology optimization. Engineering Structures, *322*, 119194, doi:https://doi.org/10.1016/j.engstruct.2024.119194.

Jeong, H., Batuwatta-Gamage, C., Bai, J., Xie, Y. M., Rathnayaka, C., et al. (2023). A complete physics-informed neural network-based framework for structural topology optimization. Computer Methods in Applied Mechanics Engineering, *417*, 116401, doi:https://doi.org/10.1016/j.cma.2023.116401.

Kumar, P. (2023). TOPress: a MATLAB implementation for topology optimization of structures subjected to design-dependent pressure loads. Structural and Multidisciplinary Optimization, *66*(4), 97, doi:https://doi.org/10.1007/s00158-023-03533-9.

Kumar, P., Frouws, J. S., & Langelaar, M. (2020). Topology optimization of fluidic pressure-loaded structures and compliant mechanisms using the Darcy method. Structural and Multidisciplinary Optimization, *61*(4), 1637-1655, doi:https://doi.org/10.1007/s00158-019-02442-0.

Le, C., Norato, J., Bruns, T., Ha, C., & Tortorelli, D. (2010). Stress-based topology optimization for continua. Structural and Multidisciplinary Optimization, *41*(4), 605-620, doi:https://doi.org/10.1007/s00158-009-0440-y.

Li, W., Bazant, M. Z., & Zhu, J. (2021). A physics-guided neural network framework for elastic plates: Comparison of governing equations-based and energy-based approaches. Computer Methods in Applied Mechanics Engineering, *383*, 113933, doi:https://doi.org/10.1016/j.cma.2021.113933.

Liu, K., & Tovar, A. (2014). An efficient 3D topology optimization code written in Matlab. Structural and Multidisciplinary Optimization, *50*(6), 1175-1196, doi:https://doi.org/10.1007/s00158-014-1107-x.

Liu, Z., Wang, Y., Vaidya, S., Ruehle, F., Halverson, J., et al. (2024). Kan: Kolmogorov-arnold networks. preprint arXiv:2404.19756, doi:https://doi.org/10.48550/arXiv.2404.19756.

Lu, L., Pestourie, R., Yao, W., Wang, Z., Verdugo, F., et al. (2021). Physics-informed neural networks with hard constraints for inverse design. SIAM Journal on Scientific Computing, *43*(6), B1105-B1132, doi:https://doi.org/10.1137/21M1397908.

Luo, Y., & Bao, J. (2019). A material-field series-expansion method for topology optimization of continuum structures. Computers and Structures, *225*, 106122, doi:https://doi.org/10.1016/j.compstruc.2019.106122.

Qiu, Q., Zhu, T., Gong, H., Chen, L., & Ning, H. (2024). Relu-kan: New kolmogorov-arnold networks that only need matrix addition, dot multiplication, and relu. preprint arXiv:2406.02075, doi:https://doi.org/10.48550/arXiv.2406.02075.

Raissi, M., Perdikaris, P., & Karniadakis, G. E. (2019). Physics-informed neural networks: A deep learning framework for solving forward and inverse problems involving nonlinear partial differential equations. Journal of Computational Physics, *378*, 686-707, doi:https://doi.org/10.1016/j.jcp.2018.10.045.

Rawat, S., & Shen, M. (2018). A novel topology design approach using an integrated deep learning network architecture. arXiv preprint arXiv:1808.02334, doi:https://doi.org/10.48550/arXiv.1808.02334.

Sanu, S. M., Aragon, A. M., & Bessa, M. A. (2024). Neural topology optimization: the good, the bad, and the ugly. arXiv preprint arXiv:2407.13954, doi:https://doi.org/10.48550/arXiv.2407.13954.

Saxena, A., & Ananthasuresh, G. (2000). On an optimal property of compliant topologies. Structural and Multidisciplinary Optimization, *19*(1), 36-49, doi:https://doi.org/10.1007/s001580050084.

Schmidt-Hieber, J. (2021). The Kolmogorov–Arnold representation theorem revisited. Neural Networks, *137*, 119-126, doi:https://doi.org/10.1016/j.neunet.2021.01.020.

Shin, S., Shin, D., & Kang, N. (2023). Topology optimization via machine learning and deep learning: a review. Journal of Computational Design and Engineering, *10*(4), 1736-1766, doi:https://doi.org/10.1093/jcde/qwad072.

Sigmund, O. (2001). A 99 line topology optimization code written in Matlab. Structural and Multidisciplinary Optimization, *21*(2), 120-127, doi:https://doi.org/10.1007/s001580050176.

Sigmund, O. (2022). On benchmarking and good scientific practise in topology optimization. Structural and Multidisciplinary Optimization, *65*(11), 315, doi:https://doi.org/10.1007/s00158-022-03427-2.

Singh, N., Kumar, P., & Saxena, A. (2025). Normalized Field Product Approach: A Parameter‐Free Density Evaluation Method for Close‐To‐Binary Solutions in Topology Optimization With Embedded Length Scale. International Journal for Numerical Methods in Engineering, *126*(7), e7673.

Sivapuram, R., & Picelli, R. (2018). Topology optimization of binary structures using integer linear programming. Finite Elements in Analysis and Design, *139*, 49-61.

So, C. C., & Yung, S. P. Higher-order-ReLU-KANs (HRKANs) for solving physics-informed neural networks (PINNs) more accurately, robustly and faster. In *2025 IEEE World AI IoT Congress (AIIoT), 2025* (pp. 1035-1042): IEEE. doi:https://doi.org/10.48550/arXiv.2409.14248.

Sohail, S. (2024). On training of kolmogorov-arnold networks. arXiv preprint arXiv:.05296, doi:https://doi.org/10.48550/arXiv.2411.05296.

Sosnovik, I., & Oseledets, I. (2019). Neural networks for topology optimization. Russian Journal of Numerical Analysis Mathematical and Modelling, *34*(4), 215-223, doi:https://doi.org/10.1515/rnam-2019-0018.

Takezawa, A., Nishiwaki, S., & Kitamura, M. (2010). Shape and topology optimization based on the phase field method and sensitivity analysis. Journal of Computational Physics, *229*(7), 2697-2718, doi:https://doi.org/10.1016/j.jcp.2009.12.017.

Van Dijk, N. P., Maute, K., Langelaar, M., & Van Keulen, F. (2013). Level-set methods for structural topology optimization: a review. Structural and Multidisciplinary Optimization, *48*(3), 437-472, doi:https://doi.org/10.1007/s00158-013-0912-y.

Wang, Y., Siegel, J. W., Liu, Z., & Hou, T. Y. (2024). On the expressiveness and spectral bias of KANs. arXiv preprint arXiv:.01803.

Yang, R.-J., & Chen, C.-J. (1996). Stress-based topology optimization. Structural and Multidisciplinary Optimization, *12*(2), 98-105, doi:https://doi.org/10.1007/BF01196941.

Yin, J., Wang, H., Guo, D., & Li, S. (2022). An efficient topology optimization based on multigrid assisted reanalysis for heat transfer problem. arXiv preprint arXiv:2210.00947, doi:https://doi.org/10.48550/arXiv.2210.00947.

Zhang, Z., Li, Y., Zhou, W., Chen, X., Yao, W., et al. (2021). TONR: An exploration for a novel way combining neural network with topology optimization. Computer Methods in Applied Mechanics Engineering, *386*, 114083, doi:https://doi.org/10.1016/j.cma.2021.114083.

Zheng, W., Wang, Y., Zheng, Y., & Da, D. (2020). Efficient topology optimization based on DOF reduction and convergence acceleration methods. Advances in Engineering Software, *149*, 102890, doi:https://doi.org/10.1016/j.advengsoft.2020.102890.